\documentclass[pra,notitlepage,superscriptaddress,twocolumn,amsmath,amssymb]{revtex4-1}

\pdfoutput=1

\usepackage{bm}
\usepackage{tikz}
\usepackage{dsfont}
\usepackage{hyperref}
\hypersetup{
	colorlinks   = true,    
	urlcolor     = blue,    
	linkcolor    = blue,    
	citecolor    = red      
}

\newcommand{\tc}{TC}
\newcommand{\ds}{DS}
\newcommand{\qd}{QD}
\newcommand{\tp}{TP}
\newcommand{\znqd}[1]{$D\left(\mathbb{Z}_#1\right)$}
\newcommand{\zz}[2]{$\mathbb{Z}_#1 \boxtimes \mathbb{Z}_#2$}
\newcommand{\z}[1]{$\mathbb{Z}_#1$}
\newcommand{\cbox}[2]{\vcenter{\hbox{\includegraphics[width=#1em]{#2}}}}
\newcommand{\dbox}[3]{\vcenter{\hbox{\includegraphics[width=#1em,height=#2em]{#3}}}}

\newcommand{\ket}[1]{\vert#1\rangle}
\newcommand{\bra}[1]{\langle#1\vert}

\begin{document}

\title{Study of anyon condensation and topological phase transitions
from a $\mathbb{Z}_4$ topological phase using Projected Entangled Pair
States}

\author{Mohsin Iqbal}
\affiliation{JARA Institute for Quantum Information, 
RWTH Aachen University, 52056 Aachen, Germany}
\affiliation{Max-Planck-Institute of Quantum Optics, 
Hans-Kopfermann-Str.\ 1, 85748 Garching, Germany}
\author{Kasper Duivenvoorden}
\affiliation{JARA Institute for Quantum Information, 
RWTH Aachen University, 52056 Aachen, Germany}
\author{Norbert Schuch}
\affiliation{Max-Planck-Institute of Quantum Optics, 
Hans-Kopfermann-Str.\ 1, 85748 Garching, Germany}

\begin{abstract}

We use Projected Entangled Pair States (PEPS) to study topological quantum
phase transitions. The local description of topological order in the PEPS
formalism allows us to set up order parameters which measure condensation
and deconfinement of anyons, and serve as a substitute for conventional
order parameters.  We apply these order parameters, together with
anyon-anyon correlation functions and some further probes, to characterize
topological phases and phase transitions within a family of models based
on a $\mathbb Z_4$ symmetry, which contains $\mathbb Z_4$ quantum double,
toric code, double semion, and trivial phases.  We find a diverse phase
diagram which exhibits a variety of different phase transitions of both
first and second order which we comprehensively characterize, including
direct transitions between the toric code and the double semion phase.
\end{abstract} 

\maketitle

\section{Introduction}

Topological phases are exotic states of matter with a range of remarkable
properties~\cite{wen:book}: They display ordering which cannot be
identified by any kind of local order parameter and requires global
entanglement properties to characterize it; they exibit strange
excitations with unconventional statistics, termed anyons; and the physics
at their edges displays anomalies which cannot exist in genuinely
one-dimensional systems and thus must be backed up by the non-trivial
order in the bulk. There has been steadily growing interest in the physics
of these systems, both in order to obtain a full understanding of
all possible phases of matter, to use their exotic properties in the design of
novel materials, and to utilize tham a as a way to reliably store quantum
information and quantum computations, exploiting the fact that the absence
of local order parameters also makes their ground space insensitive to any
kind of noise.

At the same time, the reasons which makes these system suitable for novel
applications such as quantum memories also make them hard to understand,
for instance when trying to classify and identify topological phases and
study the nature of transitions between them. Landau theory, which uses
local order parameters quantifying the breaking of symmetries to classify
phases and describe transitions between them, cannot be applied here due
to the lack of local order parameters.  Rather, phases are distinguished
by the way in which their entanglement organizes, and by the topological
-- this is, non-local -- nature of their excitations.  Formally, one can
understand the relation of certain topological phases through a formalism
called \emph{anyon condensation},  which provides a way to derive one
topological theory from another one by removing parts of the anyons by the
mechanism of condensation and confinement~\cite{bais2009condensate}.
While on a formal level, condensation should give rise to an order
parameter, in analogy to other Bose-condensed systems such as the BCS
state, it is unclear how to formally define such an order parameter in a
way which would allow to use it to characterize topological phase transitions in
a way analogous to Landau theory.

Tensor Network States, and in particular Projected Entangled Pair States
(PEPS)~\cite{verstraete:mbc-peps,verstraete2004renormalization,orus:tn-review},
constitute a framework for the local description of correlated quantum
systems based on their entanglement structure. It is based on a local
tensor which carries both physical and entanglement degrees of freedom, and
encodes the way in which these degrees of freedom are intertwined.  This
makes PEPS both a powerful numerical
framework~\cite{verstraete2004renormalization,orus:tn-review}, and a
versatile toolbox for the analytical study on strongly correlated systems;
in particular, they are capable of exactly describing a wide range of
topological ordered
systems~\cite{verstraete:comp-power-of-peps,buerschaper:stringnet-peps,gu:stringnet-peps}.
In the last years, it has been successively understood where this
remarkable expressive power of PEPS originates -- this is, how it can be that
topological order, a non-trivial global ordering in the systems'
entanglement, can be encoded locally in the PEPS description: The global
entanglement ordering is encoded in local entanglement symmetries of
the PEPS tensor, this is, symmetries imposing a non-trivial structure on
the entanglement degrees of freedom.  These symmetries ultimately build up
all the topological information locally, such as topological sectors, excitations,
their fusion and
statistics~\cite{2010peps,buerschaper:twisted-injectivity,214mpoinject,bultinck:mpo-anyons}.
Recently, this formalism has been applied to study the behavior of
topological excitations across phase transitions, and signatures of
condensation and confinement had been
identified~\cite{haegeman2015shadows}. Subsequently, this has been used to
show how to construct order parameters for measuring condensation and
deconfinement within PEPS wavefunctions, which in turn have allowed to build a
mathematical formalism for understand and relating different topological phases within the
PEPS framework through anyon
condensation~\cite{duivenvoorden:anyon-condensation}.
The core insight has been
that the order parameters measuring the behavior of anyons within any
topological phase are in correspondence to order parameters which classify
the ``entanglement phase'' of the holographic boundary state which
captures the entanglement properties (and in particular the entanglement
spectrum) of the system. 

In this paper, we apply the framework for anyon condensation within the
PEPS formalism to perform an extensive numerical study of topological
phases and phase transitions within a rich family of tensor network
models. The family is based on PEPS tensors with a $\mathbb Z_4$ symmetry,
with the \znqd4 quantum double model as the fixed point, and
correspondingly $16$ types of anyons. Within this framework, we find a rich
phase diagram including two different types of $\mathbb Z_2$ topological phases,
the Toric Code model and the Double Semion model, which are both
obtained through anyon condensation from the \znqd4 model, as well as
trivial phases. The phase diagram we find exhibits transitions between all
these phases, including direct transitions between the TC and DS model
which cannot be described by anyon condensation. 

Based on the understanding of anyon condensation within tensor networks, we
introduce order parameters for anyon condensation and deconfinement, as well
as ways of extracting correlation functions between pairs of anyons, which
allows us to characterize the different topological phases and the
transitions between them in terms of order parameters and correlation
functions, and to study their behavior and critical scaling in the
vicinity of phase transitions. Using these probes, we comprehensively explore the phase
diagram of the above family of topological models, and find a rich 
structure exhibiting both first and second order phase transitions, with
transition lying in a number of different universality classes, including a
class of transitions with continuously varying critical exponents.  In
particular, we find that the transition between Double Semion and Toric
Code phases can be both first and second order, and can exhibit different
critical exponents, depending on the interpolating path chosen.

The paper is organized as follows. In Sec.~\ref{sec:TNS-QD}, we
introduce tensor networks and the concepts relevant to topological order,
anyonic excitations, and anyon order parameters. We then introduce the
family of topological models based on \z{4}-invariant tensors which we
study in this work: We start in Sec.~\ref{sec:topo_ph} by
introducing the corresponding fixed point models and discussing their symmetry
patterns, and proceed in Sec.~\ref{sec:Topo_Phase_Trans} to describe the ways in
which we generate a family of models containing all those fixed points.
In Sec.~\ref{sec:num-methods}, we give a detailed account of the different
numerical probes we use for studying the different phases and their
transitions, and discuss how they can be used to identify the nature of a
transition.  In Sec.~\ref{sec:num-results}, we then apply these tools to
map out the phase diagram of the families introduced in 
Sec.~\ref{sec:Topo_Phase_Trans} and comprehensively study the transitions between
them.  The results are summarized in Sec.~\ref{sec:conclusions}.

\section{Tensor network formalism and \znqd{N} quantum doubles}\label{sec:TNS-QD}

In this section, we introduce PEPS and give an overview of the relevant
concepts and notions which appear in the description of topological phases
with tensor networks, with a special focus on \znqd{N} quantum doubles and
phases obtained from there by anyon condensation. We will show how
these ideas are connected and later we will use them for our study of
topological phases and phase transitions.

\subsection{Tensor network descriptions}\label{subsec:TNDesc}

\begin{figure}[t]
	\centering
	\includegraphics[width=24em]{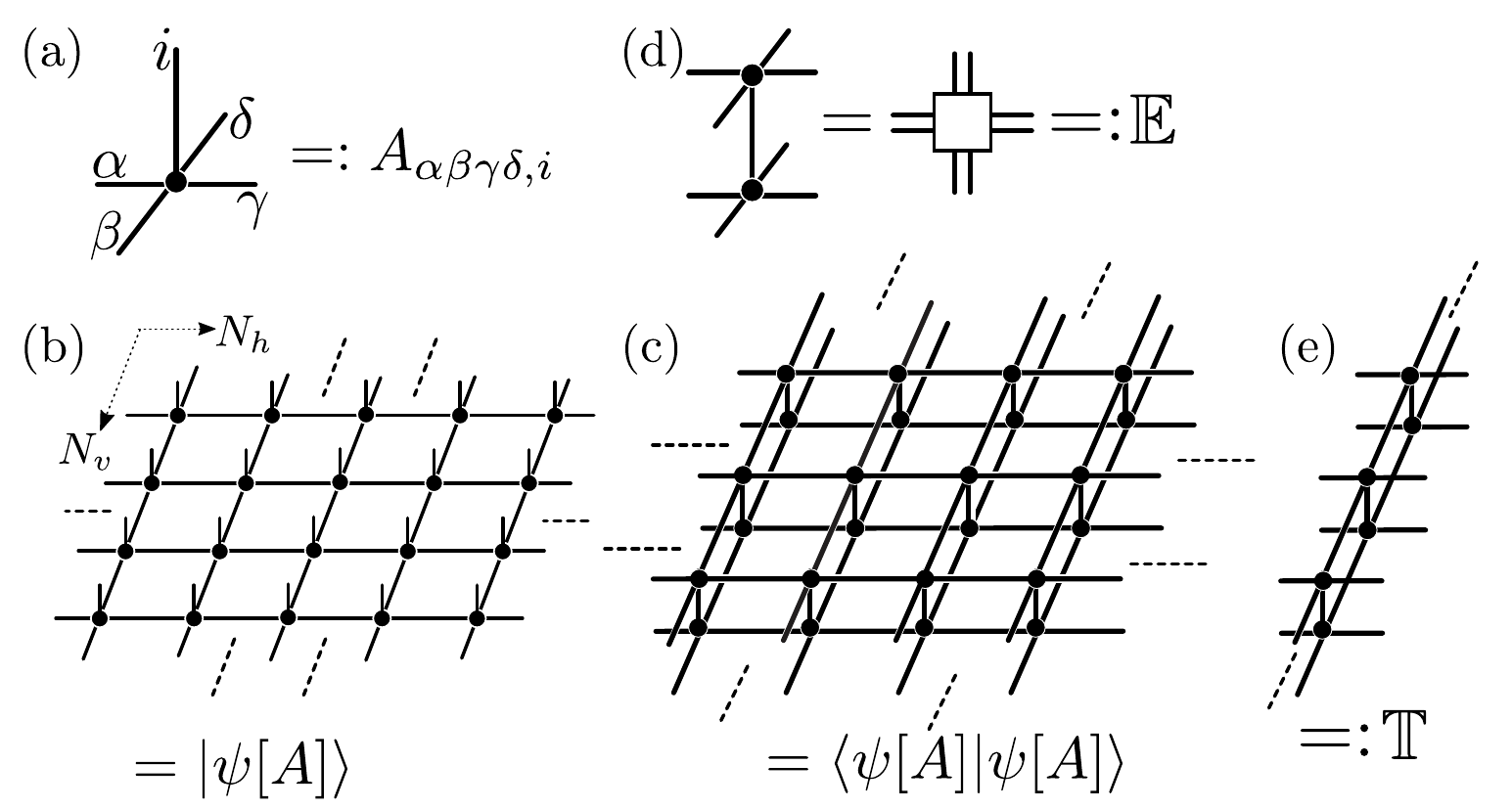}
	\caption{(a) Graphical notation of an on-site tensor $A$. (b)
Tensor network representation of a many-body wavefunction, where connected
legs are being contracted. (c) Tensor network representation of the
wavefunction norm. (d) The on-site transfer operator $\mathbb{E}$. It is
obtained by contracting the physical indices of $A$ and its conjugate. 
(e) The transfer operator $\mathbb{T}$ obtained by blocking $\mathbb{E}$
tensors in one direction.} \label{fig:toolbox}
\end{figure}

PEPS describe a many body wavefunction $|\psi\rangle$ in terms of an
on-site tensor $A_{\alpha\beta\gamma\delta;i}$ (Fig.~\ref{fig:toolbox}a).
Here, the Roman letter $i$ denotes the physical degree of freedom at a
given site, while the Greek letters denote the so-called virtual indices
or entanglement degrees of freedom
used to build the wavefunction. The many-body wavefunction is then
constructed by arranging the tensors in a 2D grid, as shown in
Fig.~\ref{fig:toolbox}b, and contracting the connected virtual indices,
i.e., identifying and summing them. This construction applies both 
to systems with periodic boundaries, to infinite planes, and to
semi-infinite cylinders.

Given a description of a many body wavefunction in terms of a local tensor
(Fig. \ref{fig:toolbox}b), the computation of the expectation value of
local observables and the norm of wavefunctions can be reduced to a tensor
network contraction problem, see Fig.~\ref{fig:toolbox}c. We will now
systematically discuss the objects which appear in the course of this
contraction.  The contraction of the physical indices of a single tensor
$A$ and its conjugate leads to the tensor
$\mathbb{E}:=\sum_{i}{A_{\alpha\beta\gamma\delta;i}A^{*}_{\alpha'\beta'\gamma'\delta';i}
}$, Fig.~\ref{fig:toolbox}d, which we refer to as the \emph{on-site
transfer operator}.  It can be interpreted as a map between virtual spaces
in the ket and bra layer. By contracting one row or column of $\mathbb E$,
we arrive at the \emph{transfer operator} $\mathbb{T}$, Fig.
\ref{fig:toolbox}e. It mediates any kind of order and correlations in the
system, and it will be an object of fundamental importance for our
studies.  More details on transfer operators will be discussed in
Sec.~\ref{subsec:bp_sop}.

\subsection{Virtual symmetries and \znqd{N} \qd}

Symmetries of the virtual indices, or briefly virtual symmetries, play a
crucial role in the characterization of the topological order carried by a
PEPS wavefunction~\cite{2010peps,buerschaper:twisted-injectivity,214mpoinject}. Virtual
symmetries are characterized by the invariance under an action on the
virtual indices of tensor $A$. More precisely, $A$ is called $G$-invariant
if one can pull through the action of $G$ on the virtual legs of $A$,
\begin{equation}\label{eq:string-action} \cbox{2.8}{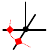} =
\cbox{2.8}{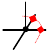} , \end{equation}
where the red squares represent a unitary group action $U_g$, $g\in G$, with $U_g$
a (faithful) representation of $G$, this is,
\[
\sum_{\alpha'\beta'}(U_g)_{\alpha\alpha'}(U_g)_{\beta\beta'}
    A_{\alpha'\beta'\gamma\delta;i}
= 
\sum_{\gamma'\delta'}(U_g)_{\gamma'\gamma}(U_g)_{\delta'\delta}
    A_{\alpha\beta\gamma'\delta';i}.
\]
We can alternatively consider $A$ as a map from virtual to the physical
space, in which case \eqref{eq:string-action} states that $A$ is supported
on the $G$-invariant subspace, i.e., the subspace invariant under the
group action $U_g\otimes U_g\otimes \bar U_g\otimes \bar U_g$. If this map
is moreover injective on the $G$-invariant subspace, it is called
$G$-injective; and if it is an isometry, $G$-isometric.

A specific case of interest is given by $\mathbb Z_N$-isometric tensors,
since they naturally provide a description of the ground space of the
so-called \znqd{N} quantum double (QD) models of
Kitaev~\cite{kitaev2003fault}. A \z{N}-isometric tensor can be constructed
by averaging over the group action of \z{N}:
\begin{equation}\label{eq:znisometric}
A_{\text{\znqd{N}}} = \sum_{g=0}^{N-1}{X^{g} \otimes X^{g} \otimes X^{g\dagger} \otimes X^{g\dagger}}
\end{equation}
up to normalization (which we typically omit in the following), where
$X^g=\sum_{i=0}^{N-1}{|i+g\rangle \langle i|}$, with
$X:=\sum_{i=0}^{N-1}{|i+1\rangle \langle i|}$ the generator of the regular
representation of \z{N}.

\subsection{Anyonic excitations of the \znqd{N} \qd}

The anyonic excitations of the \znqd{N} \qd\ are labeled by group elements
(fluxes) and irreducible representations (irreps) of
\z{N}~\cite{kitaev2003fault}. Starting from a tensor network description
of the ground state (anyonic vacuum), such an anyonic excitation can be
constructed by placing a string of $X_g$'s along the virtual degrees of
freedom of the PEPS, and terminating it with an endpoint which transforms
like an irrep $\alpha=e^{2\pi i k/N}$ of $\mathbb Z_N$, for instance  
$Z_\alpha := \sum_{i=0}^{N-1}{\alpha^i}|i\rangle\langle i|$ or 
$XZ_\alpha$~\cite{2010peps} (both of which we will use later on). A state
with one such anyon thus looks like 
\begin{equation}\label{eq:anyon_excit}
\cbox{12.5}{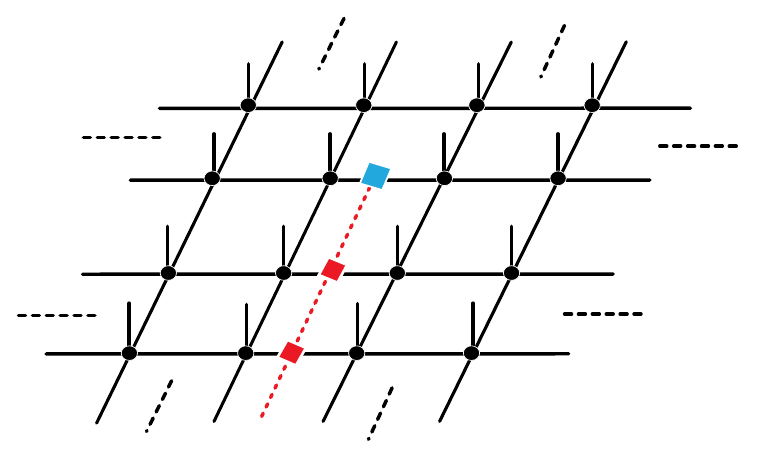} = |g,\alpha \rangle, 
\end{equation}
where the red squares on the red string denote the group action (flux)
$X^g$ and the blue square represents the irrep endpoint (charge), which in
our case will be always chosen as $Z_\alpha$.  By virtue of
Eq.~\eqref{eq:string-action}, the (red) string of $X^g$ can be freely
deformed, and thus, only its endpoint can be observed and forms an
excitation.  Excitations with trivial $\alpha=1$ are termed \emph{fluxes},
while those without string ($g=0$) are termed \emph{charges}, and
excitations with both non-trivial $g$ and $\alpha$ are called
\emph{dyons}. The statistics of the excitations is determined by the
commutation relation of fluxes and charges, $\alpha^g$.  In a slight abuse
of notation, we will denote both the state with an anyon and the anyon
itself by $|g,\alpha\rangle$. Also, we will denote the vacuum state by
$|0,1\rangle$.

In principle, such excitations need to come in pairs or tuples with
trivial total flux and charge: On a torus, this is evident since strings
cannot just terminate and the overall irrep must be trivial; with open
boundaries, the boundary conditions must compensate the topological
quantum numbers of the anyons and thus cannot be properly normalized with
respect to the vacuum unless the total flux and charge are trivial. 
However, since we can place the anyons very far from each other, and we
assume the system to be gapped (i.e., have exponentially decaying
correlations), there is no interaction between the anyons, and their
behavior (in particular the order parameters introduced in the next
section) factorize.  It is most convenient to construct a state with
trivial total flux and charge by (i) using only anyons of the same type --
this way, we can uniquely assign a value of the order parameter to a
single anyon (in principle, only their product is known) -- and (ii)
arranging them along a column at large distance, which allows to map the
problem to boundary phases, as explained in Sec.~\ref{subsec:bp_sop}.  Let
us add that since the numerical methods employed in this work explicitly
break any symmetry related to long-range correlations between pairs of
distant anyons, any such expectation value can be evaluated for a single
anyon (cf.~Sec.~\ref{subsec:order-params}).

\subsection{Condensation and deconfinement fractions}\label{subsec:condconf}

We have seen that anyonic excitations in a $\mathbb Z_N$-isometric PEPS
can be modeled by strings of group actions with dressed endpoints.
Remarkably, while describing pairs of physical anyons, these string
operators act solely on the entanglement degrees of freedom  in the tensor
network. They continue to describe anyonic excitations when deforming
the tensor away from the $\mathbb Z_N$-isometric point by acting with an
invertible operation on the physical indices (thus preserving $\mathbb
Z_N$-invariance), without the need to ``fatten'' the strings as it were
the case for the physical string operator which create anyon pairs. However, if the
deformation becomes too large, the topological order in the model must
eventually change, even though the deformation still does not affect the
string operators which describe the anyons of the \znqd{N} \qd\ phase.
However, as different topological phases are
characterized by different anyonic excitations, there must be a change in
the way in which these virtual strings correspond to physical excitations.
The mechanisms underlying these phase transitions is formed by the closely
related phenomena of \emph{anyon condensation} and \emph{anyon confinement}
\cite{bais2009condensate}. In the following, we briefly review the two
concepts in the context of tensor networks, with a particular focus on how
to use the tensor network formalism to derive order parameters for
topological phase transitions.  A rigorous and detailed account of these
ideas in the context of tensor networks is given in Ref.~\cite{duivenvoorden:anyon-condensation}.

\begin{figure}[t]
	\centering
	\includegraphics[width=25em]{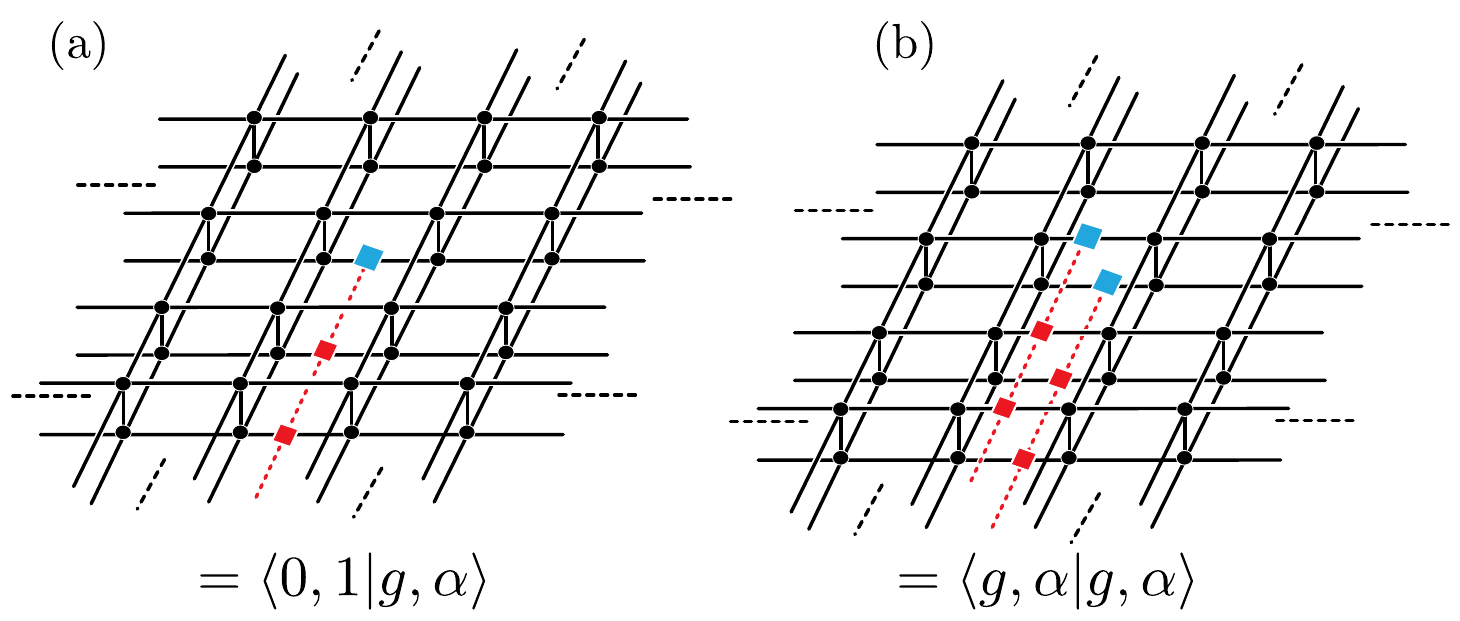}
	\caption{Graphical representation of wavefunctions overlaps in the
thermodynamic limit. (a) Condensate fraction of $|g,\alpha\rangle$.  (b)
Deconfinement fraction of $|g,\alpha\rangle$.
} \label{fig:CondConf}
\end{figure}

Condensation describes the process where an anyonic excitation becomes
part of the (new) vacuum, this is, it does no longer describe an
excitation in a different topological sector than the vacuum.  More
precisely, we say an anyon $|g,\alpha\rangle$ has been condensed to the
vacuum $|0,1\rangle$ if
\begin{equation}
\label{eq:condfrac}
\langle 0,1 | g,\alpha \rangle  \neq 0\ ,
\end{equation}
with the notation defined in Fig.~\ref{fig:CondConf}a [cf.~also
Eq.~\eqref{eq:anyon_excit}]. 

The phenomenon dual to condensation is confinement of anyons; indeed,
condensation of an anyon implies confinement of all anyons which braid
non-trivially with it~\cite{bais2009condensate,duivenvoorden:anyon-condensation}. When
separating a pair of confined anyons, their normalization goes
exponentially to zero, and thus, an isolated anyon which is confined has
norm zero.  Within PEPS, we thus say that an anyon is confined if 
\begin{equation} 
\label{eq:conffrac}
\langle g,\alpha | g,\alpha \rangle = 0\ ,
\end{equation}
again with the notation of Fig.~\ref{fig:CondConf}b.

Condensation and confinement of anyons play a crucial role in
characterizing the relation of different topological phase.  Within a
given $\mathbb Z_N$ symmetry, any topological phase can be understood as
being obtained from the $D(\mathbb Z_N)$ QD through condensation of
specific anyons, and is thus characterized by its distinct anyon
condensation and confinement
pattern~\cite{bais2009condensate,duivenvoorden:anyon-condensation}. In order to further study
the transition between different topological phases, we can generalize the
above criteria Eqs.~\eqref{eq:condfrac} and \eqref{eq:conffrac} for
condensation and confinement to order parameters, measuring the
\[
\mbox{``condensate fraction''\ } \langle 0,1|g,\alpha\rangle\;, 
\quad\mbox{Fig.~\ref{fig:CondConf}a}\ ,
\]
 and the
\[
\mbox{``deconfinement fraction''\ } \langle g,\alpha|g,\alpha\rangle\;, 
\quad\mbox{Fig.~\ref{fig:CondConf}b}\ ,
\]
respectively.  These constitute non-local order parameters for topological
phases, which can therefore be used as probes to characterize topological
phase transitions and their universal behavior, in complete analogy to
conventional order parameters in Landau theory.  The most general quantity
of interest which we will consider, encompassing both condensation and
deconfinement fractions, will thus be the overlaps $\langle
g',\alpha'|g,\alpha\rangle$ of wavefunctions describing two arbitary
anyons, with condensate and deconfinement fractions as special
cases. Note that as of the discussion in the preceding section, $\langle
g',\alpha'|g,\alpha\rangle$ is only determined up to a phase, and we will
choose it to be positive.

\subsection{Boundary phases and string order parameters}\label{subsec:bp_sop}

\begin{figure}[!t]
	\centering
	\includegraphics[width=26em]{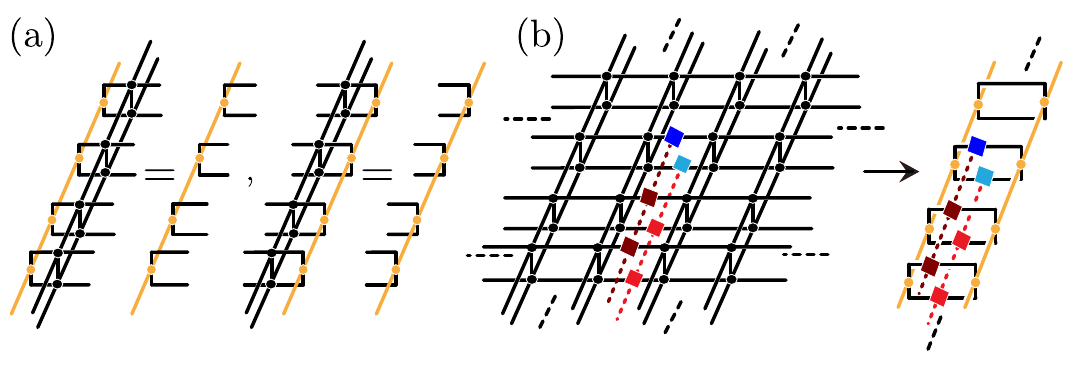}
	\caption{(a) Left and right fixed point of the transfer operator
(we assume the largest eigenvalue to be $1$); a possible Matrix Product
State (MPS) structure, found e.g.\ for fixed point models and used in the
numerics, is indicated by the yellow bonds.  (b) Mapping of anyonic
wavefunctions overlaps, Fig.~\ref{fig:CondConf}b, to the expectation value of
string order parameters.} 
\label{fig:num_2}
\end{figure}

The behavior of condensate and deconfinement fractions is closely related
to the phases encountered at the boundary of the system, this is, in the
fixed point of the transfer operator $\mathbb T$~\cite{duivenvoorden:anyon-condensation}.  To
understand this relation, consider a left and right fixed point $(l|$ and
$|r)$ of $\mathbb T$, see Fig~\ref{fig:num_2}a. Then, any such order
parameter for the behavior of anyons can be mapped to the evaluation of
the corresponding anyonic string operator -- this is, a string
$(X^g,X^{g'})\equiv(g,g')$
and irreps $(Z_\alpha,Z_{\alpha'})\equiv(\alpha,\alpha')$ in ket and bra
indices -- inbetween $(l|$ and $|r)$,
see Fig.~\ref{fig:num_2}b. (If the fixed point is not unique, $(l|$ and
$|r)$ have to match; in our case, this is easily taken care of since
$\mathbb T$ is hermitian.) Since $\mathbb T$ inherits a $G\times G$
symmetry from the tensor, an irrep $(\alpha,\alpha')$ serves as an order
parameter which detects breaking of the symmetry.  The numerical methods
we use will always choose to break symmetries when possible, which
explains why it is sufficient to consider a single anyon (rather than a
pair of anyonic operators, which would also detect long-range order
without explicit symmetry breaking.)  Similarly, strings $(g,g')$ create
domain walls in the case of a broken symmetry, and thus detect unbroken
symmetries.  Finally, combinations of irreps $(\alpha,\alpha')$ and
strings $(g,g')$ form string order parameters, which detect non-trivial
symmetry protected (SPT) phases of unbroken symmetries.

We thus see that the behavior of anyonic order parameters is in direct
correspondence to the phase of the fixed points of the transfer operator
under its symmetry $G\times G$ -- this is, its symmetry breaking pattern
and possibly SPT order -- as long as these fixed points are described by
short-range correlated states (as is expected in gapped phases).
Specifically, in the case of the fixed point tensors which we discuss in
the next section,  the fixed points are themselves MPS, and we will be
able to determine their SPT order and the behavior of the anyonic order
parameters analytically.

\section{Topological phases: Fixed points}\label{sec:topo_ph}

We now describe the tensor network constructions for the renormalization
group (RG) fixed points of topological phases which can be realized by
\z{4}-invariant tensors. The form of the local tensor which we use for
describing the RG fixed point of a topological phase is motivated by the
desired symmetry properties of its transfer operator fixed points, and
from the given tensors,  we explicitly derive the fixed points of the
transfer operator. Furthermore, we discuss the condensation and
confinement pattern of anyons in each of those topological phases. 

\begin{figure}[!t]
	\centering
	\includegraphics[width=18em]{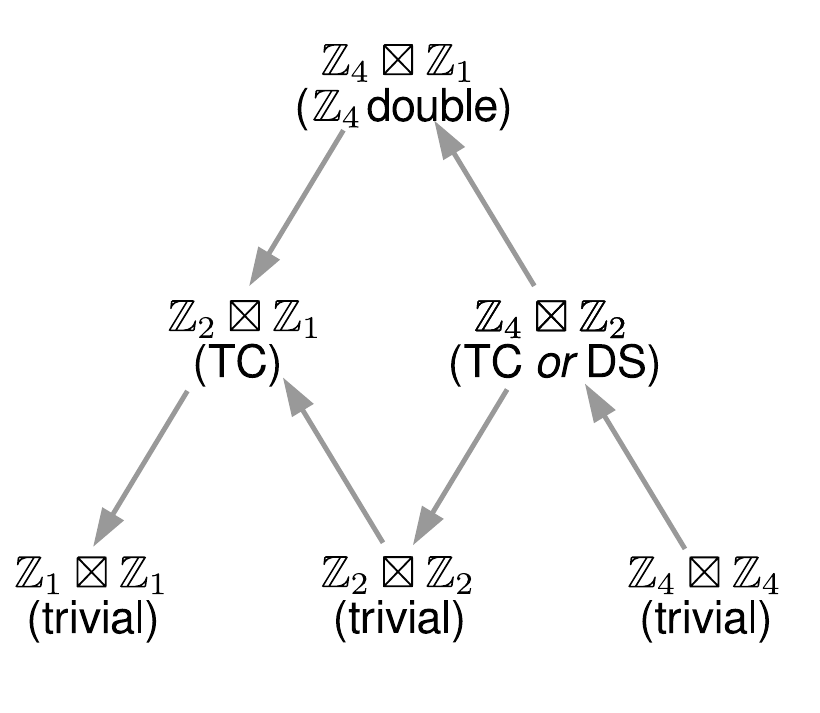}
	\caption{Patterns of symmetry breaking in the fixed points of the
transfer operator for $\mathbb Z_4$--invariant tensors, where TC/DS
denote Toric Code/Double Semion phases. Arrows indicate phase transitions
breaking a $\mathbb Z_2$ symmetry.}
\label{fig:diagLayout}
\end{figure}

The different topological phases which can be realized in the case of
\z{N}-invariant tensors, and the corresponding symmetry breaking patterns,
have been studied in Ref.~\cite{duivenvoorden:anyon-condensation}. For the case of
\z{4}-invariant tensors, the different phases and the symmetry of their
transfer operator fixed points are given in Fig. \ref{fig:diagLayout},
where the notation \zz{i}{j}$\cong \mathbb{Z}_i \times \mathbb{Z}_j$
denotes a diagonal $\mathbb Z_i$ symmetry (this is, acting identically on
ket and bra) and an off-diagonal $\mathbb Z_j$ symmetry (this is, acting
on one index alone).  For example, \zz{4}{2} is generated by the
diagonal element $(X,X)$ and the off-diagonal element $(\mathds{1},X^2)$.

\subsection{\znqd{4} quantum double}\label{subsec:z4qd}

The \znqd{4} quantum double (\qd) is a topological model which can be
realized by placing $4$-level states $\{0,1,2,3\}$ on the oriented edges of a
square lattice and enforcing Gauss' law at each vertex,
\begin{equation}\label{eq:arrow_constraint}
\cbox{3.5}{figures/gauss_arrows}\ ,\
r_1+r_2-r_3-r_4=0 \ (\text{mod} \ 4)\ ;
\end{equation}
the \znqd4 QD is then obtained as the uniform superposition over all such
configurations.
A tensor network description of \znqd{4} \qd\ is given by
a \z{4}-isometric tensor as defined in Eq. \eqref{eq:znisometric}. We begin
by writing down the on-site tensor (Fig.~\ref{fig:toolbox}a) for the
ground state of \znqd{4} \qd\ as a Matrix Product Operator (MPO):
\begin{equation}\label{eq:z4_mpo_1}
\cbox{3}{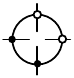}
\ ,\ \ 
\cbox{3}{figures/qd_tensor}=\delta_{ab}X^{a}
\end{equation}
and $a,b \in \left\{0,1,2,3 \right\}$. Here, the inner indices $r_i$
jointly correspond to the physical index of the on-site tensor, and the
outside indices $v_i$ to the virtual indices.  Empty circles denotes the
hermitian conjugate (with respect to the physical+virtual indices).

The relation of the construction Eq.~\eqref{eq:z4_mpo_1} with the \znqd{4}
double model can be understood by considering $X$ in its diagonal (irrep) basis.
Then, any entry of the on-site tensor is non-zero
precisely if Eq.~\eqref{eq:arrow_constraint} is satisfied, i.e.,
\begin{equation}\label{eq:z4_mpo_2}
\cbox{3}{figures/qd_mpo}=
\begin{cases} 1 & v_1+v_2-v_3-v_4=0 \ (\text{mod}\ 4), r_i=v_i
\\ 0 &  \text{otherwise}\end{cases},
\end{equation}
where $r_i , v_i \in \left\lbrace 0,1,2,3\right\rbrace$, which after
contraction yields precisely the equal weight superposition of all
configurations satisfying the $\mathbb Z_4$ Gauss law
[Eq.~(\ref{eq:arrow_constraint})], and thus the
wavefunction of the \znqd{4} model.  

Now, let us construct the fixed points of the transfer operator.  First,
note that with the properly chosen normalization factor, the MPO in
\eqref{eq:z4_mpo_1}, as a map from outside to inside, is a hermitian
projector. (Here and in what follows, all such statements will be up to
normalization.) Thus, the on-site transfer operator,
Fig.~\ref{fig:toolbox}d, is again of the form \eqref{eq:z4_mpo_1}. The
fixed points of the transfer operator and their symmetry properties can be
deduced by using the following property of MPO tensors:
\begin{equation}\label{eq:delta_property}
\cbox{3}{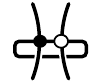}=\cbox{3}{figures/productproperty_right}\
,
\end{equation}
where the object on the right denotes a $\delta$-tensor (i.e., it is $1$
if all indices are equal, and $0$ otherwise).  Using
Eq.~\eqref{eq:delta_property}, we can write the transfer operator as
\begin{equation}\label{eq:transferop_rings_1}
\cbox{10}{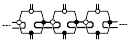}=\cbox{10}{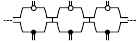}
\end{equation}
where the top/bottom indices denote both the ket and bra indices [this is,
the inside and outside indices of Eq.~\eqref{eq:z4_mpo_1}].
The $\delta$-tensors between adjacent sites force the indices in the loops
to be equal, and thus the transfer operator can be written as a sum over
four product operators,
\begin{equation}\label{eq:transferop_rings_2}
\mathbb{T}=\sum_{g=0}^{3}{\left(X^g \otimes X^{g\dag} \right)^{\otimes
N_v}}\ . \end{equation}
The fixed points of this transfer operator are then 
the following four product states:
\begin{equation}\label{eq:z4fxpoints} \left( X^g \right)^{ \otimes N_v },
\ \text{where}\ g \in \{ 0,1,2,3\}\ ,  \end{equation}
since $\mathrm{tr}(X^gX^{h\dagger})=\delta_{gh}$.
Each of the fixed points of the transfer operator in \eqref{eq:z4fxpoints}
breaks the \zz{4}{4} symmetry of the transfer operator down to \zz{4}{1},
since they are invariant under the diagonal action
$(X,X):X^g\mapsto X X^g X^\dagger$ on the bra and ket index,
but get cyclically permuted by the off-diagonal action
$(X,\openone):X^g\mapsto XX^g = X^{g+1}$. It should be noted that this
symmetry breaking structure directly originates from the block structure
of the on-site transfer operator [which is the same as
Eq.~(\ref{eq:z4_mpo_1})],
where each of the ket/bra leg pairs is simultaneously in the state $X^g$
for $g=0,\dots,3$.

The behavior of operators which act trivially (up to a phase factor) on
the fixed points of the transfer operator can be understood by their
action on the local tensors describing the fixed point space, which are
again of the form $\delta_{ab}X^a$, as in Eq.~(\ref{eq:z4_mpo_1}). For
each symmetry broken fixed point, the fixed point MPO
(cf.~Fig.~\ref{fig:num_2}a) is thus of the form
\begin{equation}\label{eq:z4_symmprop}
\cbox{14}{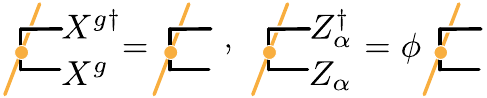},
\end{equation}
with a trivial MPO bond dimension (yellow) $D_{\mathrm{MPO}}=1$, where
$\phi$ is an additional phase factor whose exact value depend on the
specific fixed point as well as the charge label $\alpha$. 
We now see that any symmetry action $\left(X^{g_1}, X^{g_2}\right)$ with
$g_1=g_2$, as well as any irrep action 
$\left(Z_{\alpha_1}, Z_{\alpha_2}\right)$ with $\alpha_1=\alpha_2$, leave
the fixed point invariant (up to a phase).  On the other hand, acting with
either $g_1\ne g_2$ or $\alpha_1\ne \alpha_2$ on the fixed point yields
a locally orthogonal tensor, as can be either computed explicitly from
Eq.~\eqref{eq:z4fxpoints} or inferred from the commutation relations with
Eq.~\eqref{eq:z4_symmprop}. We thus arrive at 
$$
\langle g',\alpha' | g,\alpha \rangle = \begin{cases} 
1 & \mbox{if\ }g'=g\mbox{\ and\ } \alpha'=\alpha
\\0 & \text{otherwise}\end{cases}
\ .
$$
This is, no anyon is condensed, and all anyons are deconfined, and we have
a model with the full $\mathbb Z_4$ anyon content, as expected.

\subsection{Toric Codes}

Next, we will discuss how to construct Toric Codes (TC), i.e., \znqd2
double models, starting from the tensor of the \znqd4 model, while keeping
the
$\mathbb Z_4$ symmetry of the tensor. We will do so by acting with certain
projections on the physical indices, which reduce or enhance the \zz{4}{1}
symmetry of the transfer operator fixed points of the \znqd4 model to
\zz{4}{2} and \zz{2}{1}, respectively, and which yield two  Toric Codes
which are related by an electric-magnetic
duality~\cite{buerschaper2013electric}.

\subsubsection{\zz{4}{2} Toric Code}

Let us first show that by acting with the  projector
$(\mathds{1}+X^2)^{\otimes 4}$ on the local tensors of the \znqd{4}
\qd, we obtain a tensor for the RG fixed point of the \tc\
phase where the fixed points of the transfer operator have a \zz{4}{2}
symmetry. The new tensor is obtained from \eqref{eq:z4_mpo_1} as
\begin{equation}\label{eq:Z2Z4tcmpo}
\cbox{3}{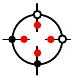}\ ,\
\cbox{3}{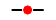}=\mathds{1}+X^2.
\end{equation}
The action of the projections, denoted by red dots, on the black ring
(i.e., the \znqd4 tensor) gives two independent blocks, and the resulting
tensor can be written as an MPO with bond dimension two:
\begin{equation}\label{eq:Z2Z4tcmpo2}
\cbox{3}{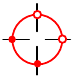} = \cbox{3}{figures/Z2Z4TC_mpo}
\ ,\ \ 
\cbox{3}{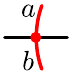}=\delta_{ab}X^a\left(\mathds{1}+X^2\right),
\end{equation}
and $a,b\in \{0,1\}$. The reason why Eq.~\eqref{eq:Z2Z4tcmpo2} provides a 
tensor network description for the Toric Code can be
understood from the equivalence
\begin{equation}\label{eq:Z2Z4tcmpo3}
\cbox{3}{figures/red_tensor} = \delta_{ab}\left(|+\rangle \langle+|\otimes \sigma_x^a \right)
\end{equation}
where $a,b\in \{0,1\}$ and $\sigma_x$ is the generator of \z{2}. Up to the
$\ket{+}\bra{+}$, it is thus exactly of the same form as
\eqref{eq:z4_mpo_1}, but with the underlying group $\mathbb Z_2$, and thus
describes a \znqd2 model (i.e., the Toric Code).  The construction of
Eq.~(\ref{eq:Z2Z4tcmpo2}) can therefore be viewed as a \znqd{2}
model, tensored with an ancilla qubit in the $\ket+$ state.

The on-site transfer operator of this model again satisfies the delta
relation of Eq.~\eqref{eq:delta_property} (now with only two possible values), and
each of the two blocks in Eq.~\eqref{eq:Z2Z4tcmpo2} can be identified with
a symmetry broken fixed point of the transfer operator, which are thus of
the form
\begin{equation}
\left( \mathds{1}+X^2\right)^{\otimes N_v},\ \left( X+X^3\right)^{\otimes
N_v}\ .
\end{equation}
Each of the fixed points is invariant under the action of $\left( X,X
\right)$ and
$\left(\mathds{1},X^2\right)$, while $(\openone,X)$ transforms between
them. We thus find that the Toric Code model at hand has a \zz42
symmetry in the fixed point of the transfer operator, and we thus
henceforth call it the \zz42 Toric Code.

Graphically, the actions which leave fixed points invariant can be summarized as follows
\begin{equation}\label{eq:z4xz2_symmprop}
\cbox{18}{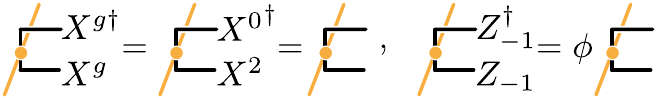},
\end{equation}
where $\phi$ is a phase factor with a value which depends on the fixed
point, while all other actions yields locally orthogonal tensors. 
We can thus summarize the condensation and confinement pattern of anyons
at the RG fixed point of \zz{4}{2} \tc\ phase as 
$$
\langle g',\alpha' | g,\alpha \rangle = 
\begin{cases} 
1 & g'=g\ (\text{mod }2),\ \alpha' = \alpha = \pm 1
\\ 0 & \text{otherwise}, 
\end{cases}				
$$
This implies that the anyon $|2,1\rangle$ is condensed, while the
anyons with $\alpha=\pm i$ have become confined, giving rise to a TC with
anyons $|1,1\rangle$, $|0,-1\rangle$, and $|1,-1\rangle$ (the
fermion).

\subsubsection{\zz{2}{1} Toric Code}

A second construction for a \tc\ phase in the framework of \z{4}-invariant
tensors is obtained by a dual projection.  We will see that this
projection, in contrast to the \zz{4}{2} \tc, reduces the symmetry of
fixed points of the transfer operator to \zz{2}{1}, and the fixed point
space of the transfer operator is spanned by eight symmetry broken fixed
points. The projection action on the physical indices is given by $(\mathds{1}+Z^2)^{\otimes
4}+(\mathds{1}-Z^2)^{\otimes 4}$, and, acting on 
the \znqd{4} \qd\ tensor, generates an MPO with eight blocks,
\begin{equation}
\label{eq:TC-z2z1-tensor}
\cbox{3}{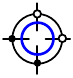}\ ,\ 
\cbox{3}{figures/Z2TC_tensor}=\delta_{ab}\left(\mathds{1}+(-1)^{a}Z^2\right),
\end{equation}
$a,b \in \left\lbrace0,1\right\rbrace$, and $Z:=Z_1$. Together with the
index of the other ring, Eq.~(\ref{eq:z4_mpo_1}), the bond dimension
around the circle is $8$.

It can be checked directly that (\ref{eq:TC-z2z1-tensor}) is again
a hermitian projector, and that the on-site transfer
operator satisfies a delta relation as in Eq.~\eqref{eq:delta_property}.
Thus, we find that the fixed points of the transfer operator
are
\begin{equation}\label{eq:z2tc-fxpoints}
{\left(X^a\left( \mathds{1}+(-1)^bZ^2\right)\right)}^{\otimes N_v},
\end{equation}
where $a \in \{0,1,2,3\}$ and $b\in \{0,1\}$. 
They are invariant under $(X^2,X^2)$, while any other symmetry action
results in a permutation action on the fixed points. We thus
find that the symmetry of the fixed point space is given by \zz21.
The set of all symmetries of the fixed point tensor is thus given by 
\begin{equation}\label{eq:z2_symmprop}
\cbox{18}{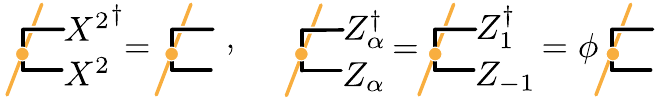}\ ,
\end{equation}
while all other actions with $X^g$ or $Z_\alpha$ give rise to orthogonal
tensors. The overlap of anyonic wavefunctions is thus
$$
\langle g',\alpha' | g,\alpha \rangle = \begin{cases} 
1 & g'=g=0 \text{ or } 2, \ \alpha' =\pm\alpha
\\ 0 & \text{otherwise}\ . \end{cases}
$$
This implies that $|0,-1\rangle$ is condensed, while the anyons with
$g=1$ and $g=3$ are confined.  The anyons of the TC are thus give by
$|0,i\rangle$, $|2,1\rangle$, and $|2,i\rangle$ (the fermion).

\subsection{\zz{4}{2} Double semion model}

A model closely related to the \tc\ is the double semion (\ds) model. It
also corresponds to a $\mathbb Z_2$ loop model, but is twisted with 
a $3$-cocycle which assigns an amplitude $(-1)^\ell$ to loop configurations with
$\ell$ loops.  It has no tensor network description in terms of
\z{2}-invariant tensors, and its description either requires
\z{4}-invariance~\cite{iqbal2014semionic}, or tensors which are injective
with respect to MPO-symmetries~\cite{214mpoinject}. As we
will see, the fixed points of the transfer operator of the \ds\ model also
have a \zz{4}{2} symmetry \cite{duivenvoorden:anyon-condensation}, but they
realize a different phase under that symmetry as compared to the \zz{4}{2}
\tc, which is characterized by a non-zero string order parameter (i.e., an
SPT phase), corresponding to the fact that it is
obtained by condensing a dyon (a composite charge-flux particle) in the
\znqd4 model.
The local tensor of the \ds\ model can be constructed by applying the
following MPO projector (green) on the \znqd{4} \qd,
\begin{equation}\label{eq:dsmpo}
\cbox{3}{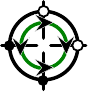}\ ,\
\cbox{3}{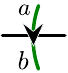}={\left(X^2\right)}^{a}{\left(Z^2\right)}^{a+b},
\end{equation}
with $a\in\{0,1\}$; arrows in the ring point in the direction of index
$b$.  
While it can be shown that this PEPS can be transformed to the \qd\ model
by local unitaries~\cite{duivenvoorden:anyon-condensation}, we will instead directly derive the
fixed points of the transfer operator and show that they exhibit the
symmetry pattern required for the \ds\ phase.  The composition of the
\znqd{4} \qd\ (black ring) and  \ds\ (green ring) projector can
be simplified to
\begin{equation}
\label{eq:ds_mpo_full}
X^{2a+c}Z^{2(a+b)}\mbox{\ \ with\ }a,b\in\{0,1\}\ ,
\end{equation}
where the index $c$ is identical on the top and bottom leg. (Here, we have
used a redundancy in the description which allows to restrict
$c=0,\dots,3$ to $c=0,1$, and removed phases which cancel out between
adjacent tensors.)

As in the previous cases, the tensor \eqref{eq:dsmpo} is a hermitian
projector. Connecting two tensors (in the form~\eqref{eq:ds_mpo_full}]
as required for the transfer operator, cf.~Eq.~\eqref{eq:delta_property},
yet again gives rise to a $\delta$ tensor for all three indices $a$, $b$,
and $c$. The fixed points are thus given by MPOs with tensors of the form
\begin{equation}\label{eq:dsmpstensor}
\cbox{3}{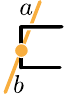}=\cbox{3}{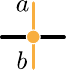}=
X^{2a+c}Z^{2(a+b)}
\end{equation}
where $c=0,1$ labels the two fixed points, and the green line is
the MPO index for the fixed point MPO, with $a,b=0,1$.

From (\ref{eq:dsmpstensor}), it can be seen that the symmetry actions which
leave the tensor invariant are
\begin{equation}\label{eq:ds_symmprop}
\cbox{21}{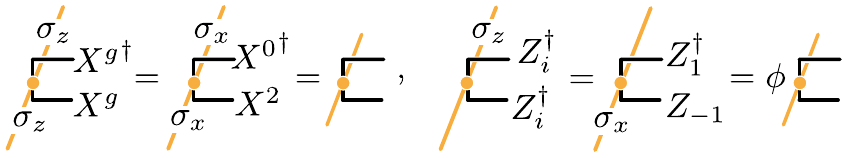}\ ,
\end{equation}
while $(X,\openone)$ permutes the two fixed points, and $(Z_1,Z_i)$ maps it
to a locally orthogonal tensor. It follows that the fixed point has \zz42
symmetry, which is however not detected by local order parameters, but
requires the use of string order parameters. The non-vanishing string
order parameters can be read off Eq.~\eqref{eq:ds_symmprop}: On the one
hand, this is $(Z_1,Z_{-1})$ with a string of $(\openone,X^2)$ going
downwards, and on the other hand, $(XZ_i,XZ_i)$ with a string of $(X,X)$
going downwards.  It is crucial to notice that here, we have to fix the
direction in which the string is pointing (since the virtual actions of
the irreps are asymetric), and in the latter case, the endpoint has to be
``dressed'' by using $XZ_i$ as an irrep. This effectively moves the
virtual action $\sigma_z$ to the lower leg; otherwise, the string order
parameter would vanish even though it were allowed for topological
reasons, i.e., from the combination of group action and
irrep it carries.  

This leads us to adapt the definition of the anyons,
cf.~Eq.~\eqref{eq:anyon_excit}, used in this work by choosing irrep
endpoints $XZ_{i}$ and $XZ_{-i}$ for anyons with a string $X$ or
$X^3$, i.e., $|1,\pm i\rangle$ and $|3,\pm i\rangle$, while using
$Z_\alpha$ as endpoints for all other anyons.
Importantly, the results on the former phases (and the trivial phases
described later) still hold with these modified anyons. In fact, the
only other phase where these anyons are not confined -- in which case
string operators involving them yield zero for topological reasons, i.e.,
solely due to the choice of group element and irrep of the
string~\cite{duivenvoorden:anyon-condensation} -- is the \znqd4 model, in
which case it can be easily checked that the modified anyons are still
uncondensed and deconfined.

The presence of string order with respect to the order parameters for the
unbroken symmetry implies that the fixed point states describe a
non-trivial SPT phase with symmetry \zz42.  Using
Eq.~(\ref{eq:ds_symmprop}), we find
$$
\langle g',\alpha' | g,\alpha \rangle = \begin{cases} 
1 & g,g'\text{\ even},\ \alpha'=i^{g-g'}\alpha=\pm 1
\\
1 & g,g'\text{\ odd},\ \alpha'=i^{g-g'}\alpha=\pm i
\\
0 & \text{otherwise}. \end{cases}				
$$
This shows that in the DS phase, only the dyon $|2,-1\rangle$ is condensed.
The anyons with $g$ even and $\alpha=\pm i$, as well as $g$ odd and
$\alpha=\pm1$, are condensed.  The semions of the model are $|1,i\rangle$
and $|1,-i\rangle$, and the boson is $|0,-1\rangle\equiv|2,1\rangle$.

\subsection{Topologically trivial phases}

Let us finally discuss how to obtain topologically trivial phases (TP) from the
\znqd4\ model. We will find that there are three different ways of
obtaining such models, distinguished by their condensation/confinement
pattern.

\subsubsection{\zz{4}{4} trivial phase}

The first trivial phase is obtained by applying a projector $P^{\otimes
4}$, 
\begin{equation}
\label{eq:Z4Z4-trivial-proj}
P={\left(\mathds{1}+X+X^2+X^3\right)}\ ,
\end{equation} to the physical indices of
the \znqd{4} \qd, which projects each of the indices indiviually on the
trivial irrep.  Clearly, this commutes with the symmetry action of the
\znqd4\ projector, Eq.~(\ref{eq:z4_mpo_1}), and thus, the on-site transfer
operator itself is of the form $P^{\otimes 4}$, which in turn implies that
the fixed point of the transfer operator is unique, and of the form
$P^{\otimes N_v}$ as well.  The model thus has \zz44 symmetry at the boundary. 
A manifestation of this  symmetry is the condensation of all the flux
anyons, and correspondingly confinement of all charged particles. The
condensation and confinement properties of anyons in the \zz{4}{4} \tp\ can be
summarized as
$$
\langle g',\alpha' | g,\alpha \rangle = \begin{cases} 
1 & \alpha'= \alpha = 1
\\0 & \text{otherwise}. \end{cases}				
$$

\subsubsection{\zz{2}{2} trivial phase}

The second trivial phase is obtained by applying an MPO projector (brown) to the
\znqd4\ projector as follows:
\begin{equation}
\label{eq:z22-triv-proj}
\cbox{3.2}{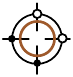}
\ , \
\cbox{3.2}{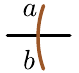}=\delta_{ab}\left(\mathds{1}+X^2\right)\left(\mathds{1}+(-1)^{a}Z^2\right),
\end{equation}
where $a,b=\in \{0,1\}$. Combining it with the \znqd4\ projector with
tensor $X^c$, we find that $c$ can be restricted to $c\in\{0,1\}$, and
\eqref{eq:z22-triv-proj} yields again a hermitian projector. Contraction
of adjacent on-site transfer operators yields yet again a Kronecker delta
on the two loop variables, leading to four fixed points of the form
\begin{equation}
\big((X^c+X^{c+2})(\openone+(-1)^aZ^2)\big)^{\otimes N_v}\ .
\end{equation}
It is immediate to see that these are invariant under $(\openone,X^2)$,
and are permuted by $(\openone,X)$ and $(X,X)$ (acting on $c$ and $a$,
respectively). Thus, this phase has symmetry \zz22.  Overall, this
yields
$$
\langle g',\alpha' | g,\alpha \rangle = \begin{cases} 
1 & g,g'\in\{0,2\},\ \alpha', \alpha \in \{1,-1\}
\\0 & \text{otherwise}\end{cases}\ :
$$
Anyons with $g=0,2$ and $\alpha=\pm1$ are condensed, while anyons with
$g=1,3$ or $\alpha=\pm i$ are confined, making the anyon content trivial.

\subsubsection{\zz{1}{1} trivial phase}

The last trivial phase is obtained by acting on the \znqd4\ with an MPO
projector with bond dimension $4$,  
\begin{equation} 
\label{eq:z11-triv-proj}
\cbox{3}{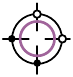}\ , \
\cbox{3}{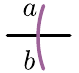}=\delta_{ab}
\sum_{j=0}^{3}(i^a)^j{Z^j}\propto\delta_{ab}\ket{3-a}\bra{3-a}\ ,
\end{equation}
where $a,b \in \left\{0,1,2,3\right\}$. The effective MPO which is
obtained by composing black and violet rings has bond dimension $16$, with
tensor elements $\ket{a}\bra{c}$. It is thus clearly a hermitian
projector, and has fixed points of the form $(\ket{c}\bra{c'})^{\otimes
N_v}$ which clearly break all symmetries, giving rise to a \zz11 symmetry
of the fixed point. 
The phase is trivial with all charges condensed and thus all anyons with
non-trivial flux (including dyons) confined, 
$$
\langle g',\alpha' | g,\alpha \rangle = \begin{cases} 
1 & g=g'=0
\\0 & \text{otherwise}. \end{cases}
$$
	

\section{Topological phases: Interpolations and transitions}\label{sec:Topo_Phase_Trans}

The fixed point tensors discussed in the preceding section all share the
$\mathbb Z_4$-invariance as a common feature. It is therefore suggestive
to try to build smooth interpolations within these tensors, e.g.\ by
starting from the $\mathbb Z_4$-isometric \znqd4 tensor and deforming it
smoothly towards some other fixed point model.  As long as this
deformation is reversible (as will be the case here), it corresponds to a
smooth deformation of the parent Hamiltonian~\cite{schuch:mps-phases}, and
thus forms a tool to study the phase diagram of the corresponding model.
Since, as we have discussed in the previous section, the different phases
are characterized by different symmetries at the boundary and thus
different anyon condensation patterns, this will give rise to topological
phase transitions which are (potentially) driven by anyon condensation.

In the following, we will discuss a number of interpolations which allow
us to study all the phases in Fig.~\ref{fig:diagLayout}.  The
interpolations are obtained using two different recipes:
The first approach aims to interpolate between the on-site tensors of the
respective models in an as local as possible way; we will refer to this
approach as \emph{local filtering}. The second approach is based on
interpolating between the on-site transfer operators rather than the
tensors, which however can be shown to correspond to a smooth path of
tensors due to some positivity condition; we will refer to it as
\emph{direct interpolation of transfer operator}.

The interpolations described in the following, and in particular the
nature of the phase transitions, will be studied numerically
in Sections~\ref{sec:num-methods} and \ref{sec:num-results}.

\subsection{Three-parameter family with all topological phases
\label{subsec:3par-family-def}}

We start by describing a three-parameter family which exhibits all
four topological phases in Fig.~\ref{fig:diagLayout}, together with the
\zz22 trivial phase. To start with, we define the following three
one-parameter interpolations.

First,
\begin{equation}\label{eq:Z4_Z2xZ4TCPT}
\cbox{3}{figures/Z2Z4TC_mpo}\ ,\text{ where }
\cbox{3.5}{figures/Z2Z4TC_tensor}=\exp (\theta_\text{\tc} X^2 ).
\end{equation}
For $\theta_{TC}=0$, this acts trivially, while for $\theta_{TC}=\infty$,
this yields the \zz42 TC, Eq.~\eqref{eq:Z2Z4tcmpo}.  

Second,
\begin{equation}\label{eq:Z4_Z2TCPT}
\cbox{3.2}{figures/Z2TC_mpo}\:, \text{ where }
\cbox{3}{figures/Z2TC_tensor}=\delta_{ab}\exp \left( (-1)^{a}\theta_{\text{\tc},Z_2} Z^2 \right)
\end{equation}
and $a,b\in \left\{0,1\right\}$: 
For $\theta_{\mathrm{TC},Z_2}=0$, this acts trivially, while for
$\theta_{\mathrm{TC},Z_2}=\infty$, this yields the the \zz21
TC, Eq.~\eqref{eq:TC-z2z1-tensor}.  

Third,
\begin{equation}\label{eq:Z4_Z2xZ4DSPT}
\cbox{3.6}{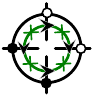}\ ,\text{ where } 
\cbox{3.4}{figures/ds_tensor_1}=
{\left(X^2\right)}^{a}{\left(Z^2\right)}^{a+b}
\end{equation}
and $\cbox{3}{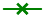}=\mathrm{diag}(\cosh
\frac{1}{2}\theta_\text{\ds}, \sinh \frac{1}{2}\theta_\text{\ds})$. At
$\theta_\text{\ds}=0$, $\cbox{3}{figures/ds_tensor_x}=\left| 0
\right\rangle\left\langle 0 \right|$, and whole ring acts trivially,
whereas for $\theta_\text{\ds}=\infty$,
$\cbox{3}{figures/ds_tensor_x}\propto\mathds{1}_2$, and the green ring
acts as the \ds\ projector, Eq.~\eqref{eq:dsmpo}.

It is straightforward to verify that all three deformations
(\ref{eq:Z4_Z2xZ4TCPT}--\ref{eq:Z4_Z2xZ4DSPT}) commute both among each
other and with the \znqd4 projector. We can thus combine them to obtain a
three-parameter tensor 
\begin{equation}\label{eq:z4phasediag}
\cbox{4.2}{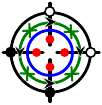} \ ,
\end{equation}
which is parametrized by $\bm{\theta} := \left(\theta_{\text{TC}},
\theta_{\text{TC},Z_2}, \theta_{\text{DS}} \right)$.  From the mutual
commutation, it follows that the limits $\bm\theta=(\infty,0,0)$,
$\bm\theta=(0,\infty,0)$, and $\bm\theta=(0,0,\infty)$ still yield
the corresponding fixed point models, and $\bm\theta=(0,0,0)$ the \znqd4\
model. Finally, it is straightforward to check that by setting any two
of the $\theta_\bullet=\infty$, we obtain the \zz22 trivial phase.

\subsection{Transitions into trivial phases
\label{subsec:interpolations_introtrivial}}

The three-parameter family of the previous section did not include all the
trivial phases present in Fig.~\ref{fig:diagLayout}. We now give three
families interpolating from each of the different $\mathbb Z_2$
topological phases to the trivial ones.

\subsubsection{\zz{4}{2} \tc\ and trivial phases}

The \zz{4}{2} \tc\ can be deformed into \zz{4}{4} or \zz{2}{2}
\tp\ through the two-parameter family
\begin{align}\label{eq:Z2xZ4TCTP_PT}
\cbox{3}{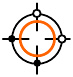}\ , \text{ where }
\cbox{3}{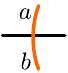} =\delta_{ab}\frac{1}{2}
\left[\cosh\theta_1\left(\mathds{1}+X^2\right) \right.+\nonumber  \\
\left.  \sinh\theta_1\left(X+X^3\right)\right]\exp\left((-1)^a\theta_2
Z^2\right), \end{align}
$a,b\in \left\{0,1\right\}$, and where the black ring represents the local
tensor of the \znqd{4} \qd. It is built such as to interpolate between the
tensors \eqref{eq:Z2Z4tcmpo}, \eqref{eq:Z4Z4-trivial-proj}, and
\eqref{eq:z22-triv-proj} of the respective fixed point
models: For $(\theta_1,\theta_2)=(0,0)$, it gives the \zz{4}{2} \tc, for
$(\theta_1,\theta_2)=(\infty,0)$  the \zz{4}{4} \tp, and for
$(\theta_1,\theta_2)=(0,\infty)$ the \zz{2}{2} \tp.

\subsubsection{\zz{2}{1} \tc\ and trivial phases}

The \zz{2}{1} \tc\ can be connected to the \zz22 \tp\ and the \zz11 \tp\
through the family of deformations \begin{align}\label{eq:Z2TCTP_PT}
\cbox{3}{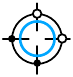}, \cbox{3}{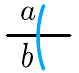}
=\delta_{ab}\frac{1}{2} \left[\cosh\theta_1 \left(\mathds{1}+(-1)^a
Z^2\right) \right.+ \nonumber \\  \left. i^a \sinh\theta_1 \left(Z+(-1)^a
Z^3\right)\right]\exp\left(\theta_2X^2/4\right) \end{align} where $a,b\in
\left\{0,1,2,3\right\}$.  Again, it interpolates between the three fixed
points \eqref{eq:TC-z2z1-tensor}, \eqref{eq:z22-triv-proj}, and
\eqref{eq:z11-triv-proj}, with 
$\left(\theta_1,\theta_2\right)=\left(0,0\right)$ the \zz21 \tc, 
$(\theta_1,\theta_2)=(\infty,0)$ the \zz11 trivial phase, and 
$(\theta_1,\theta_2)=(0,\infty)$ the \zz22 trivial phase.

\subsubsection{\zz{4}{2} \ds\ and trivial phases}

Finally, we describe an interpolation which connects the
\zz{4}{2} \ds\ and the 
\zz{2}{2} \tp\ and \zz{4}{4} \tp:  
\begin{equation}\label{eq:DSTP_PT}
\cbox{3}{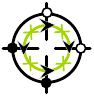}\ , \text{ where }
\cbox{3}{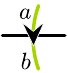}=
{\left(X^2\right)}^{a}{\left(Z^2\right)}^{a+b}\exp{\left(\theta_2
X\right)}, \end{equation}
$a=0,1$, and where
$\cbox{3}{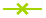}=\exp\left(\theta_1 \sigma_z\right)$
with $\sigma_z$ the Pauli $z$ matrix. Its extremal points are 
at $(\theta_1,\theta_2)=(0,0)$ the \zz42 \ds\ model, at 
$(\theta_1,\theta_2)=(0, \infty)$ the \zz{2}{2} \tp, and at
$(\theta_1,\theta_2)=(\infty, \infty)$ the \zz44 \tp. Moreover, the point
$(\theta_1,\theta_2)=(\infty, 0)$ realizes the \znqd4 model. As we will
see in Sec.~\ref{subsec:PhDiag_TCDS_TPs}, there exists a direct path
between \zz{4}{2} \ds\ and \zz{4}{4} \tp\ via a multi-critical point.

\subsection{Direct interpolation of transfer operator
\label{subsec:EFL}}

Let us now describe our second approach to constructing interpolations,
the direct interpolation of the transfer operator.  While they in
principle form a special case of local filtering operations, the idea here
is to construct an interpolation on the level of the on-site transfer
operators, rather than the tensor, which yields different interpolation
paths and in some cases seems to be more robust in retrieving direct phase
transitions between the involved phases. 

We start from two on-site tensors $A_1$ and $A_2$ which form the
RG fixed point of two distinct phases. We assume that $A_1$ and $A_2$ are
both $G$-invariant, i.e., they can be obtained by applying linear maps
$\bm P_i$ on the local tensor of the \znqd{4} \qd, i.e.,
$A_i=\bm P_iA_{\text{\znqd{4}}}$, where $A_{\text{\znqd{4}}}$ denotes the
on-site tensor of the \znqd{4} \qd. 

Now consider a direct interpolation of the on-site transfer operators
$\mathbb{E}_m=A_{m}^\dagger A_m$, 
\begin{equation}\label{eq:LinearInt2}
\mathbb{E}\left(\theta\right)=\theta \mathbb{E}_1 + (1-\theta) \mathbb{E}_2,
\end{equation}
$\theta\in[0;1]$.  Since $\mathbb E_m\ge0$, it follows that also $\mathbb
E(\theta)\ge0$, and thus $\mathbb E(\theta)=A(\theta)^\dagger A(\theta)$ for some
$A(\theta) =\bm P(\theta) A_{\text{\znqd{4}}}$; moreover, $A(\theta)$ can be
chosen continuous in $\mathbb E(\theta)$.   In the case where the $\bm P_i$
are commuting hermitian projectors, a continuous $A(\theta)$ can be
constructed through
\begin{equation}\label{eq:LinearInt3}
\begin{split}
A\left(\theta\right) & = \left[\bm{P}_{12} + \sqrt{\theta}(\bm{P}_1-\bm{P}_{12}) \right. \\
&  \left. + \sqrt{1-\theta}(\bm{P}_2-\bm{P}_{12})\right]
A_{\text{\znqd{4}}},
\end{split}
\end{equation}
where $\bm{P}_{12} := \bm{P}_1\bm{P}_2$. 

We will use this construction in three cases: (i) In
Sec.~\ref{sec:num-methods}, we will use it to interpolate between the
\zz42 DS to \zz42 TC model, i.e., $\bm P_1$ and $\bm P_2$ are the DS and
TC projectors of Eqs.~\eqref{eq:dsmpo} and \eqref{eq:Z2Z4tcmpo2}.  (ii)
Also in Sec.~\ref{sec:num-methods}, we will use it to interpolate between
the \znqd4 QD and the \zz42 TC. Here, $\bm P_1$ is trivial, and $\bm P_2$
the same as before.  (iii) Finally, in Sec.~\ref{subsec:tc21-to-ds}, we
will use it to interpolate between the \zz42 DS and the \zz21 TC model,
with $\bm P_1$ and $\bm P_2$ from Eqs.~\eqref{eq:dsmpo}
and~\eqref{eq:TC-z2z1-tensor}. In all these cases, $\bm P_1$ and $\bm P_2$
commute, as we have seen in Sec.~\ref{subsec:3par-family-def}.


\section{Numerics: Methods}\label{sec:num-methods}

We will now give an overview over the numerical tools and methods which we
will use to probe the phase diagram and in particular the phase
transitions between different topological phases. Most impartantly, these
are order parameters for condensation and deconfinement, as well as
different correlation lengths (including those corresponding to
anyon-anyon correlation functions involving string operators). We will
also discuss a few additional probes suitable to characterize phase
transitions. We will both describe the corresponding probes, and give
detailed account of how to compute them.  

In order to better illustrate how these probes can be used to characterize
phase transitions, and to show how to use them to distinguish first-order
from second-order phase transitions, we will study two specific
interpolations, namely \zz{4}{2}  \ds\ $\leftrightarrow$ \zz{4}{2} \tc\ and
\znqd{4} \qd\ $\leftrightarrow$ \zz{4}{2} \tc. Both of these
interpolations are constructed by linear interpolation of the on-site transfer
operators as explained in Sec.~\ref{subsec:EFL}.

\subsection{Order parameters}\label{subsec:order-params}

Order parameters play a fundamental role in characterizing the nature of
phase transitions, and are at the heart of Landau's theory of second order
phase transitions. Their behavior allows to identify different phases
through their symmetry breaking pattern, to distinguish first from second
order phase transitions, and to further characterize second order
transitions through their critical exponents.  While topological phases do
not exhibit local order parameters, we have seen in
Sec.~\ref{subsec:condconf} and~\ref{subsec:bp_sop} how to define order
parameters for anyon condensation and deconfinement through operators
defined on the virtual, i.e., entanglement degrees of freedom.  In the
following, we discuss how these order parameters can be measured, and how
they can be used to characterize the nature of topological phase
transitions.

\subsubsection{Computation}

\begin{figure}[t]
	\centering
	\includegraphics[width=25.4em]{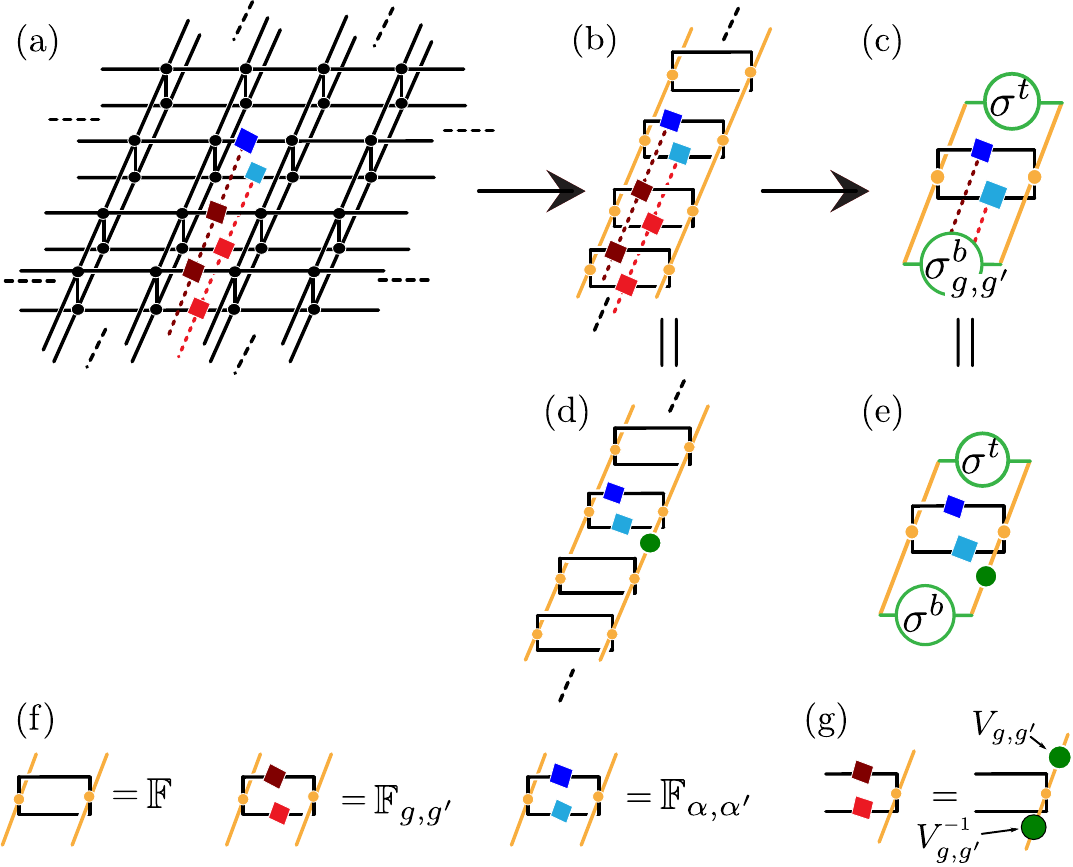}
	\caption{
Computation of anyonic order parameter $\langle g,\alpha
|g,\alpha\rangle$. The computation of the 2D order parameter (a) is
carried out by dimensional reductions to string order parameters in the 1D
iMPS boundaries (b), which are then evaluated by using the fixed points of
the \emph{channel operators} defined in (f).  In order to fix
normalization, the problem can be related (d,e) to the vacuum expectation
value by using the local representation of the symmetry string in the
boundary iMPS (g).  See text for further details.}
\label{fig:num_1}
\end{figure}

As explained in Sec. \ref{subsec:condconf}, any possible such order
parameter is given by an anyonic wavefunction
overlap $\langle g,\alpha|g',\alpha'\rangle$ (normalized by
$\langle0,1|0,1\rangle$), shown in Fig.~\ref{fig:num_1}a. Here, the
strings (red) and their endpoints (blue) correspond to group actions $g$,
$g'$ and irrep actions $\alpha$ and $\alpha'$, respectively.  To compute
this quantity, we proceed in two steps: In a first step, we approximate
the the left and right fixed points of the transfer operator by Matrix
Product Operators. This reduces the (2D) computation of $\langle
g,\alpha|g',\alpha'\rangle$ to evaluating a (1D) string-order parameter in
the left and right fixed point, Fig.~\ref{fig:num_1}b, which is then
carried out in a second step (cf.~also Sec.~\ref{subsec:bp_sop}).

The computation of the left and right fixed points $(l|$ and $|r)$ of the
transfer operator in the thermodynamic limit is carried out using a
standard infinite matrix product state (iMPS) algorithm. The basic idea is
to use an infinite translational invariant MPS ansatz with some bond
dimension $\chi$ to approximate the fixed point.  This ansatz can e.g.\ be optimized by
a fixed point method, this is, by
repeatedly applying the transfer operator to it (this increases the 
bond dimension which is truncated to $\chi$ in every step by keeping the
terms with the highest Schmidt coefficients, and is the method used in
this paper), or in the case of a hermitian transfer operator by
variationally optimizing the iMPS
tensor, e.g.\ by linearizing the problem, until convergence is reached.  A
detailed overview over different methods for finding fixed points of
transfer operators can be found in Ref.~\cite{haegeman:medley}. In all
cases, truncating the bond dimension to a finite value $\chi$ induces some
error, and therefore, an extrapolation in $\chi$ is required.  An
important point about all these methods is that they will favor symmetry
broken fixed points -- this is, whenever the fixed point is degenerate,
the method will pick a symmetry broken fixed point rather than a cat-like
state with long-range order, as the former has less correlations. As
already discussed earlier, this is the reason why we can evaluate anyonic
order parameters by considering just a single anyon (which requires
symmetry breaking to be non-zero) rather than a distant pair of anyons
(which would also detect long-range order in cat-like fixed points).

\begin{figure*}[t]
	\includegraphics[width=\textwidth]{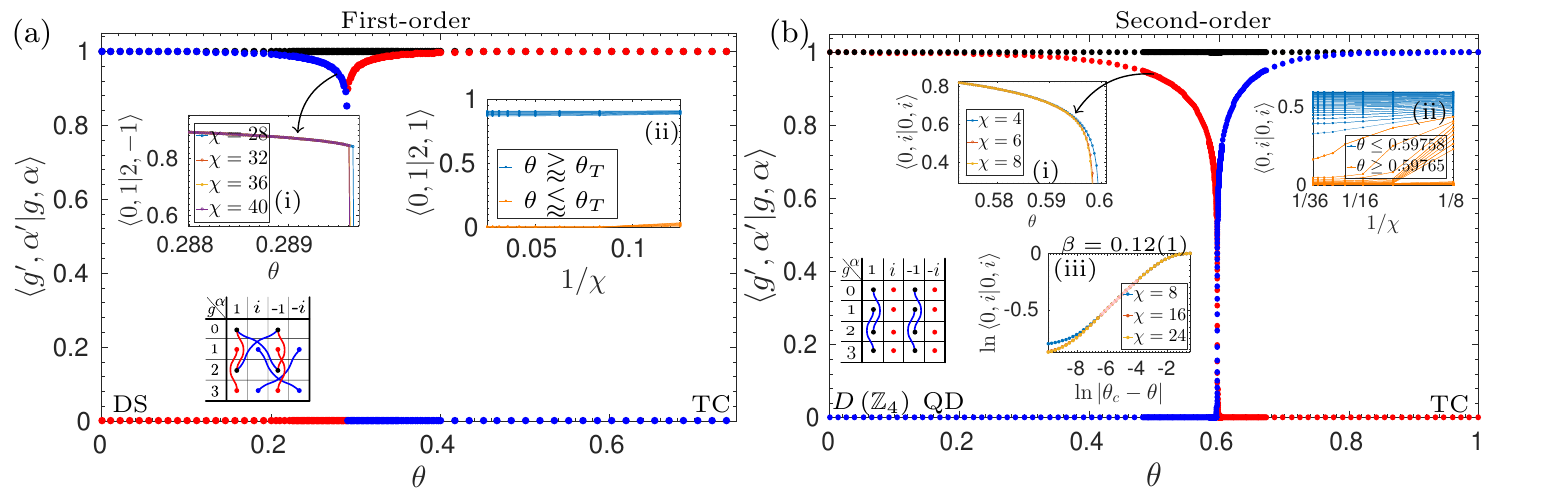}
	\caption{Analysis of phase transitions between (a) \zz{4}{2} \ds\ and \zz{4}{2}
\tc, and (b) \znqd{4} \qd\ and \zz{4}{2} \tc. The tables in the bottom
left describe the color coding: Rows/columns label flux $g$/charge
$\alpha$ of anyons, and colored dots/lines denote the color used for
plotting the order parameters $|\langle g,\alpha|g,\alpha\rangle|$ and
$|\langle g,\alpha|g',\alpha'\rangle|$, respectively. The insets (i) and
(ii) show the behavior of the order parameter indicated on the $y$-axis in
the vicinity of the phase transition, either (i) as a function of $\theta$ for
different MPO bond dimensions $\chi$, or (ii) as a function of $\chi$ for
different $\theta$.  The sharp transition in (a) indicates a
first-order transition, the smooth change in (b) a second-order
transition.  In the latter case, a critical exponent $\beta$ can be
extracted from the data, cf.\ inset (iii).
\label{fig:compare_v1}}
\end{figure*}

We have now rephrased the computation of $\langle
g,\alpha|g',\alpha'\rangle$ in terms of a string order parameter evaluated
in $(l|$ and $|r)$, as shown in Fig.~\ref{fig:num_1}b; this has to be
normalized by evaluating the same object \emph{without} the string order
parameter, i.e., $(l|r)$. In both cases, we have to evaluate an object of
the form Fig.~\ref{fig:num_1}b.  This can be carried out by considering
the transfer operators from top and bottom in Fig.~\ref{fig:num_1}b, which
we will term \emph{channel operators}, shown in Fig.~\ref{fig:num_1}f: $\mathbb{F}$ is
obtained by contracting the ``physical'' indices of the local MPO tensors
of $(l|$ and $|r)$, $\mathbb{F}_{g,g'}$ carries additional group actions
$g$ and $g'$ indicated by the red squares, and $F_{\alpha,\alpha'}$
carries the irrep actions (blue) which correspond to the desired anyon.
First, we must compare the modulus of the leading eigenvalue of $\mathbb
F_{g,g'}$ with the leading eigenvalue of $\mathbb F$: If the former is
strictly smaller, $\langle g,\alpha|g',\alpha\rangle$ will be
exponentially supressed in the length of the string, and thus be zero.
This is the case exactly if the symmetry $(g,g')$ is broken, since the
normalized leading eigenvalue determines the overlap of the original and the
symmetry-transformed fixed point per unit cell. In case the symmetry
$(g,g')$ is unbroken, we compute the largest eigenvector $\sigma_{g,g'}^b$
of $\mathbb{F}_{g,g'}$ from the bottom and the largest eigenvector
$\sigma^t:=\sigma_{0,0}^t$ of $\mathbb{F}:=\mathbb{F}_{0,0}$ from the top
by exact diagonalization.  (Note that the eigenvectors are unique, since
the fixed point iMPS will break symmetries, and $(g,g')$ are unbroken
symmetries -- a degenerate eigenvector would indicate long-range order.)
Finally, the expectation value is computed by acting on
$\mathbb{F}_{\alpha,\alpha'}$ with $\sigma^t$ and $\sigma_{g,g'}^b$ from
the top and bottom respectively as
shown in Fig. \ref{fig:num_1}c.

However, there is an important issue: We still have to fix normalization,
as the eigenvectors $\sigma_{\bullet}^\bullet$ have no well-defined
normalization.   To this end, we note that since we only consider unbroken
symmetries $(g,g')$, the symmetry is locally represented in the
fixed point iMPS $|r)$ through some action $V_{g,g'}$ and $V^{-1}_{g,g'}$,
as shown in Fig.~\ref{fig:num_1}g~\cite{sanz:mps-syms}. By suitable
rescaling $V_{g,g'}$, we can always choose $V_{g,g'}$ to be a
representation, and it will be crucial that we do so.  We now
substitute Fig.~\ref{fig:num_1}g everywhere in 
Fig.~\ref{fig:num_1}b and obtain the expression Fig.~\ref{fig:num_1}d for
the fraction $\langle g,\alpha|g',\alpha'\rangle$,
which simplifies to the expression
Fig.~\ref{fig:num_1}e with the \emph{same} $\sigma^t$ and $\sigma^b$ as in
the normalization (which has $g=g'=0$ and $\alpha=\alpha'=1$.)  In order
to fix the normalization we can thus either extract the symmetry action
$V_{g,g'}$ from the iMPS fixed
point $|r)$ through $\sigma_{g,g'}^b \propto V_{g,g'}\sigma^b$ 
and evaluate Fig.~\ref{fig:num_1}e, or we can choose a
canonical gauge for the iMPS of $|r)$ such that the symmetry action $V_{g,g'}$ is
unitary~\footnote{This is the case for both a right- or a left-canonical
gauge for $|r)$, this is, the gauge where either the right or the left
fixed point of the transfer operator of the iMPS representing $|r)$ alone
is the identity.}, and normalize $\sigma^b$ and
$\sigma^b_{g,g'}$ with a unitarily invariant norm, e.g.\
$\mathrm{tr}\,(\sigma^b)^2 = \mathrm{tr}\,(\sigma^b_{g,g'})^2=1$.

It is important to note that this choice of normalization -- which hinges
upon the choice of $V_{g,g'}$ -- is not arbitrary. First, the result is
invariant under changing the gauge of the iMPS for $|r)$ and $|l)$.
Second, normalizing $V_{g,g'}$ such as to form a representation is
necessary to obtain the same value for each anyon when aligning several
identical anyons along a column (with their strings in parallel):
Otherwise, the contribution of some of the anyons will be given
Fig.~\ref{fig:num_1}e with $V_{g,g'}$ as the green dot, while for others
it will be $V^{-1}_{g^{-1},g'^{-1}}$, which is only guaranteed to give the
same result if the $V_{g,g'}$ form a representation.

\subsubsection{Analysis}

Let us now analyze the behavior of anyonic order parameters in the case of the two
interpolations \zz{4}{2} \ds\ $\leftrightarrow$ \zz{4}{2} \tc\ and
\znqd{4} \qd\ $\leftrightarrow$ \zz{4}{2} \tc, shown in
Fig.~\ref{fig:compare_v1}a and b, respectively. Since there are $16$
anyons ($g$ and $\alpha$ can each take four possible values), there are in
total $136$ different overlaps. However, it turns out that many of these
overlaps are either zero, in which case we omit them from the figure, or
equal to each other (this can both be observed numerically and explained
from the symmetry structure of the state).  The main plots in
Fig.~\ref{fig:compare_v1}a,b show the remaining different non-zero
order parameters. The color coding in any such plot is explained in the
table in the lower left corner: The boxes in the table correspond to
different anyons (rows
label $g$, columns label $\alpha$).  The colors of the dots and lines in
the table correspond to the different non-zero order parameters: Solid dots 
represents the norm $\langle g,\alpha | g,\alpha \rangle$, and lines 
between two entries represents overlaps $\langle g,\alpha | g',\alpha'
\rangle$, where $g\neq g'$ and $\alpha \neq \alpha'$. The absence of a dot
or an edge indicates quanitites which remain zero along the whole
interpolation, this is, each such plot carries the full information on all
$|\langle g,\alpha|g',\alpha'\rangle|$.

Specifically, in Fig.~\ref{fig:compare_v1}a, the blue curve describes both
the condensate fraction of $(g=2,\alpha=-1)$ and the deconfinement
fraction of the anyon with $(g=1,\alpha=i)$, while the red curve describes the condensate
fraction of $(g=2,\alpha=1)$ and deconfinement fraction of
$(g=1,\alpha=1)$; this relates to the fact that at the DS--TC transition,
particles have to both condense/confine and uncondense/deconfine. In
Fig.~\ref{fig:compare_v1}b, the blue curve gives the condensate fraction
of $(g=2,\alpha=1)$ and the red one the deconfinement fraction of
$(g=0,\alpha=i)$: At the phase transition into the TC phase, the former
condensed and the latter becomes confined.

The order parameters in Fig.~\ref{fig:compare_v1}a and b show a different
behavior around the phase transition: In Fig.~\ref{fig:compare_v1}a, they
abruptly drop to zero, indicative of a first-order phase transition, while
in Fig.~\ref{fig:compare_v1}b, they vanish continuously, corresponding to a
second-order transition. This is confirmed by a careful analysis of the
data around the phase transition: The insets (i) in the two panels show a
magnified view of one of the order parameters (cf.~label) in the vicinity
of the phase transition for different values of $\chi$: Clearly, both
curves are well converged in $\chi$, but show a fundamentally different
behavior -- discontinuous vs.\ continuous.  This is also confirmed
by plotting the order parameter vs.~$1/\chi$ for different values of the
interpolation parameter $\theta$ around the critical point, shown in
the insets (ii): While for the
1st order transition, there is an abrupt change with a clear gap in the
value of the order parameter at the phase transition, for the 2nd order
transition its value changes smoothly, subject to a stronger
$\chi$-dependence around the transition. For the value of the phase
transition, we find $\theta_T=0.2896(5)$ for the DS--TC interpolation,
Fig.~\ref{fig:compare_v1}a, and $\theta_c=0.5976(2)$ for the \znqd4 QD--TC
interpolation, Fig.~\ref{fig:compare_v1}b.

In the case of a second-order phase transition, we can additionally
compute critical exponents $\beta$, $|\langle
g,\alpha|g',\alpha'\rangle|\propto |\theta-\theta_c|^\beta$, for the
various order parameters on both sides of the phase transition.  For the
\znqd4 QD--TC interpolation under consideration, we find $\beta=0.12(1)$
for all non-trivial order parameters, consistent with an 2D Ising
universality class, see inset (iii) in Fig.~\ref{fig:compare_v1}b.

\subsection{Correlation length}\label{sec:methods-correlation}

The other relevant quantity which can be used to characterize the behavior at the
phase transition is the scaling of correlation functions. In the case of
topologically ordered systems, this can encompass both correlation
functions of local observables as well as anyon-anyon correlation
functions, which are described by string-order type correlators.

\subsubsection{Computation}

Within the framework of PEPS, there are several different ways
to extract correlation lengths.  We will now outline three different
methods which we will make use of.

The first two are based on the fact that all correlations within PEPS are
mediated by the transfer operator (Fig. \ref{fig:toolbox}e).
Specifically, both the decay of arbitrary two-point correlations and of
anyon-anyon correlation functions are determined by the leading
eigenvalues of the transfer operator.
In order to obtain the spectrum of the transfer operator, we can follow
two routes: (1) We can use exact diagonalization of the transfer operator 
on an infinitely long cylinder with finite perimeter $N_v$ to obtain the correlation length and
then extrapolate in $N_v$ (using a fit $a\exp[-b N_v]+\xi_\infty$)
to get a reliable estimate of $\xi$ in the
thermodynamic limit.  In order to get access to both local and anyon-anyon
correlations, the exact diagonalization has to be performed on all
sectors of the transfer matrix, i.e., including the possibility of
inserting a flux $g$ ($g'$) in the ket (bra) layer when closing the
boundary, and labeling the eigenvectors by their topological charge (i.e.,
irrep label) $\alpha$ and $\alpha'$; the sector of the corresponding
correlation function is then given by the difference of ket and bra flux
and charge~\cite{schuch2013topological}.  The overall correlation length
$\xi$ can be computed from the gap in the spectrum of the transfer
operator below the largest ground space sector -- in the case of
\z{4}-invariant tensors with $4^2=16$ ground states, $\xi=-1/\ln
|\lambda_{16}/\lambda_{0}|$.  (2) Alternatively, we can compute the gap of
the transfer operator by determining its fixed point in the thermodynamic
limit using an iMPS ansatz, and then using an iMPS excitation ansatz as
proposed in \cite{pirvu2012matrix, haegeman2012variational,
haegeman2013post} to model the excitations.  In particular, the excitation
ansatz allows to also explicitly construct topologically non-trivial
excitations by attaching a flux string (=symmetry action) to the
excitation and giving it a non-trivial charge (=irrep label), and thus
allows to access the different topological
sectors~\cite{haegeman2015shadows}.

Finally, a third method to extract a correlation length is to use the
channel operator $\mathbb F$ corresponding to the fixed point of the
transfer operator (Fig. \ref{fig:num_1}f), whose spectrum can be computed
efficiently as it is system size independent. The correlation length can
again be extracted from the subleading eigenvalues of the channel
operator, as well as the the leading eigenvalues of the dressed channel
operator $\mathbb F_{g,g'}$, and additionally using irrep labels of the
eigenvectors to fully address anyonic correlations.  Note, however, that
this approach in principle only gives access to correlations along a
specific axis.

\subsubsection{Analysis}

\begin{figure}[t]
\includegraphics[width=\columnwidth]{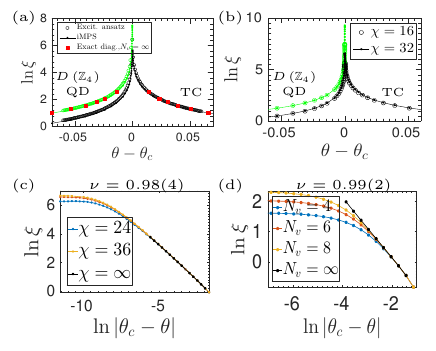}
\caption{\label{fig:compare_corr_methods}%
Comparison of data on correlation lengths for the \znqd4 QD to TC
transition, cf.~Fig.~\ref{fig:num_1}b.  \textbf{(a)} Comparison of correlation
length different methods for obtaining correlation lengths; see (b) for
the color coding.
\textbf{(b)} Correlation length for anyon-anyon correlations with flux (green) and
without flux (black); the latter includes trivial two-point correlations.
The data shows that in the \znqd4 QD phase, the transition is dominated by flux
condensation. \textbf{(c,d)} Extraction of the critical exponent from (c) the iMPS
ansatz for the boundary state, and (d)~from extrapolation of finite
cylinders.
}
\end{figure}

Let us now analyze the information obtained from the different methods.
We start by a comparison of the methods for the 2nd order \znqd4 QD to TC
transition in Fig.~\ref{fig:compare_corr_methods}.  Panel (a) compares the
results obtained from the different methods, which are in very good
agreement.  (Data from finite cylinders is only shown in the regime where
the extrapolation to $N_v\rightarrow\infty$ works reliably.)  
Fig.~\ref{fig:compare_corr_methods}b shows the correlation length
extracted from the channel operator of the iMPS fixed point, labelled by
its flux.  We see that on the left of the phase transition, the dominating
correlations are those between fluxful anyons, which indicates that the
transition from the \znqd4 QD phase is driven by condensation of fluxes
(or dyons), in accordance with what is observed in Fig.~\ref{fig:compare_v1}b
where the fluxes are condensed in the TC phase.
Fig.~\ref{fig:compare_corr_methods}c,d
finally show the extraction of the critical exponent $\nu$ from iMPS data,
extrapolated in $\chi$ (panel c), and finite cylinder data, extrapolated
in $N_v$ (panel d), which are in very good agreement.

\begin{figure}[t]
\includegraphics[width=\columnwidth]{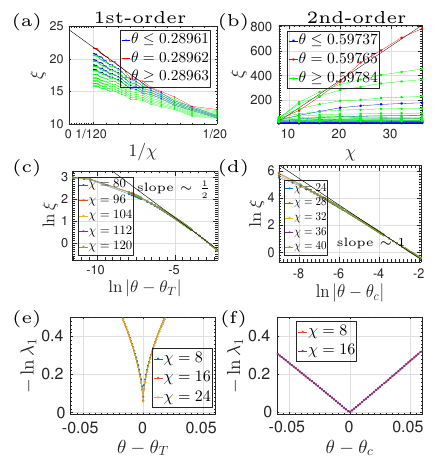}
\caption{\label{fig:compare_corrs}%
Comparison of correlations for 1st order (left column) vs.\ 2nd order
(right column) transition. \textbf{(a,b)} Data obtained iMPS, extrapolated
in $\chi$, showing convergence to constant $\xi$ (1st order) vs.\
divergent (2nd order) behavior. \textbf{(c,d)} Scaling of $\xi$ in the
vicinity of the phase transition obtained with iMPS, showing convergence
vs.\ critical scaling with $\nu=1$ as the transition point is approached.
\textbf{(e,f)} Inverse correlation lengths obtained by diagonalizing the transfer
operator using an excitation ansatz.}
\end{figure}

Fig.~\ref{fig:compare_corrs} compares the results for the 1st order
transition from DS to TC (left column) with the 2nd order transition from
\znqd4 QD to TC (right column). We find that in the 1st order case, the
correlation length obtained from iMPS converges linearly in $1/\chi$ to a
constant $\xi\approx25$ [Fig.~\ref{fig:compare_corrs}a], while in the 2nd
order case, it diverges approximately linearly in $\xi$
[Fig.~\ref{fig:compare_corrs}b]; also note that in this case the
correlation is already $\xi\sim800$ for a bond dimension $\chi=36$. 
Fig.~\ref{fig:compare_corrs}c,d shows the scaling of $\xi$ in the
vicinity of the phase transition as it is approached from the left.
While in the 1st order case, the curve leaves the initial
$|\theta_T-\theta|^{1/2}$ scaling as the transition $\theta_T$ is
approaching, and this behavior does not depend on $\chi$, the scaling in
the case of the 2nd order transitions approaches the $|\theta_c-\theta|$
scaling closer and closer to $\theta_c$ as $\chi$ is increased. Finally,
Fig.~\ref{fig:compare_corrs}e,f shows the inverse correlation length
$1/\xi=-\ln(\lambda_1)$ as extracted from the diagonalization of the
transfer operator using an excitation ansatz: We find that the eigenvalue
gap in the first-order case, while small, remains open, while it closes in
the second-order case.

\subsection{Further probes}

Our main tools to analyze phase transitions will be correlation length and
order parameters. However, there are a number of other probes which allow
us to look more closely at phase transitions, and which we describe in the
following.

\subsubsection{Fidelity susceptibility}

The overlap between ground state wavefunctions, known as fidelity, has
been pointed out as a probe in order to study quantum phase
transitions~\cite{zanardi2006ground}.  More specifically, the
\emph{fidelity susceptibility} 
\begin{equation}
\label{eq:fid-suscep-def}
\chi_F:=\lim_{\delta\rightarrow 0}
\frac{ 2(1-f(\theta,\theta+\delta))}{\delta^2}\ ,
\end{equation}
where
$f(\theta,\theta+\delta)=
\langle\psi(\theta)|\psi(\theta+\delta)\rangle/(N_h N_v)$ 
is the fidelity per site as a parameter changes from $\theta$ to
$\theta+\delta$, exhibits universal features which can be used to
characterize phase transitions~\cite{zanardi2007information}. 
An account on the behavior of fidelity per site and its computation using
iMPS algorithm is given in Appendix \ref{app:fid}. 

\begin{figure}[!t]
	\centering
	\includegraphics[width=27em]{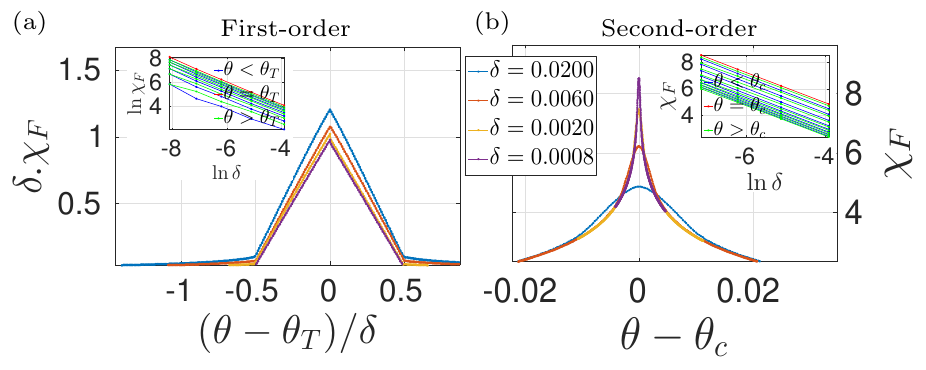}
	\caption{Fidelity susceptibility, Eq.~\eqref{eq:fid-suscep-def},
for (a) first- and (b) second-order phase transition computed using iMPS
with $\chi=32$ for different step sizes $\delta$
(cf.~Appendix~\ref{app:fid}.  In the first-order case (a), $\chi_F$
approaches a delta funtion, while in the second order case, it diverges as
$\log|\theta-\theta_c|$. The insets show the scaling of $\chi_F$ with the
step size $\delta$, which also exhibits distinct behaviors.}
\label{fig:fid_suss}
\end{figure}

In the following, we discuss the distinct features of $\chi_F$ for the two
above mentioned phase transitions, showing clearly distinct signatures for
1st and 2nd order transitions. The data is shown in
Fig.~\ref{fig:fid_suss} for different values of the step size
$\delta\rightarrow0$.  In both cases, $\chi_F$ diverges at the phase transition.  
However, the divergence is very distinct: In the case of the 1st order
transition between \zz{4}{2} \ds\ and \zz{4}{2} \tc, $\chi_F(\theta)$
converges to a delta function as $\delta\rightarrow0$, as can be seen from
the rescaled plot in Fig.~\ref{fig:fid_suss}a, which shows that
$\chi_F(\theta-\theta_T)\rightarrow\delta^{-1}\Lambda((\theta-\theta_T)/\delta)$
for a universal triangle-shaped function $\Lambda$. This is in accordance
with the expected abrupt change of the ground state wavefunction (even per
unit cell) at a first order phase transition.  For the 2nd order
transition between \znqd{4} \qd\ and \zz{4}{2} \tc, on the other hand, 
$\chi_F\sim \log|\theta-\theta_c|$ as $\delta\rightarrow0$.  (Since we
expect $\chi_F$ to scale like the structure factor for an observable
relating to the derivative of the local PEPS tensor
$A(\theta)$~\cite{zanardi2007information,schuch:rvb-kagome}, which scales
like the corresponding correlation length $\xi_A$ squared, we expect
$\xi_A$ to only diverge logarithmically with $|\theta-\theta_c|$, in
agreement the fact that we are considering a topological phase
transition.)

\subsubsection{Susceptibility
\label{subsec:methods_susceptibility}}

Order parameters measure the amount of spontaneous symmetry breaking in
the system. Further universal information about order parameters can be
extracted by computing their susceptibility, this is, the scaling of their
response to an infinitesimal field which explicitly breaks the symmetry in
the vicinity of the phase transition, 
\begin{equation} 
\chi_m(\theta) := \left.
\frac{\partial O }{\partial h}
\right|_{h=0} \ .
\end{equation}
The resulting critical exponent $\chi_m(\theta)\propto
|\theta-\theta_c|^\gamma$ allows to further characterize the phase
transition.  

In the case of topological phase transitions, $O=|\langle
g,\alpha|g',\alpha'\rangle|$ will be an order parameter for
condensation or deconfinement. The external field $h$ corresponds to
adding an infinitesimal term to the PEPS tensor which explicitly breaks
the symmetry of the transfer operator in favor of one fixed point. An
example, including more details on the computation of the susceptibility
as well as numerical results, is given in Sec.~\ref{sec:num:qd-to-42tc}
in the discussion of the \znqd4\ QD to \zz42 TC transition.

\subsubsection{Dispersion relations}

Above, we have described how to use an excitation ansatz to extract the
correlation length of a PEPS wavefunction directly in the thermodynamic
limit.  Using the same method, we can also obtain $k$-dependent
correlation functions, which give us further information about features of
the dispersion relation of the
system~\cite{zauner2015transfer}, and in particular about the mechanism
driving the topological phase transition ~\cite{haegeman2015shadows}. We
present dispersion data for the DS to \zz21 TC transition in 
Appendix~\ref{app:dispersion}.

\section{Numerics: Results}\label{sec:num-results}

In the previous section, we have discussed various numerical probes for
the study of topological phase diagrams. In the following, we will apply
these techniques to systematically explore the whole phase diagram of
\z{4}-invariant tensor network states.

The main focus of this section will be the three-parameter family
introduced in Sec.~\ref{subsec:3par-family-def} which includes \znqd4 QD,
TC, DS, and trivial phases. We start in
Sec.~\ref{subsec:results_phasediag} with summarizing the phase-diagram of
the model, and discuss the different transitions of the model in
Sec.~\ref{subsec:results_z4_z2} (transitions from the \znqd4 QD phase),
Sec.~\ref{subsec:results_tc_ds} (transitions between TC and DS), and
Sec.~\ref{subsec:results_contvar} (a phase transition with continuously
varying exponents between DS and trivial phase). Finally, in
Sec.~\ref{subsec:PhDiag_TCDS_TPs} we discuss the families interpolating
between TC/DS and trivial phases introduced in
Sec.~\ref{subsec:interpolations_introtrivial}.

\subsection{Phase diagram of \z{4}-invariant tensor network states
\label{subsec:results_phasediag}}

Our main object of interest will be the family of states defined in
Sec.~\ref{subsec:3par-family-def}, and in particular
Eq.~(\ref{eq:z4phasediag}), which by deforming a \z{4}-invariant tensor
allowed us to interpolate between the \znqd4 QD, Toric Code, Double Semion, and
trivial phases.

\begin{figure}[t]
	\includegraphics[width=23em]{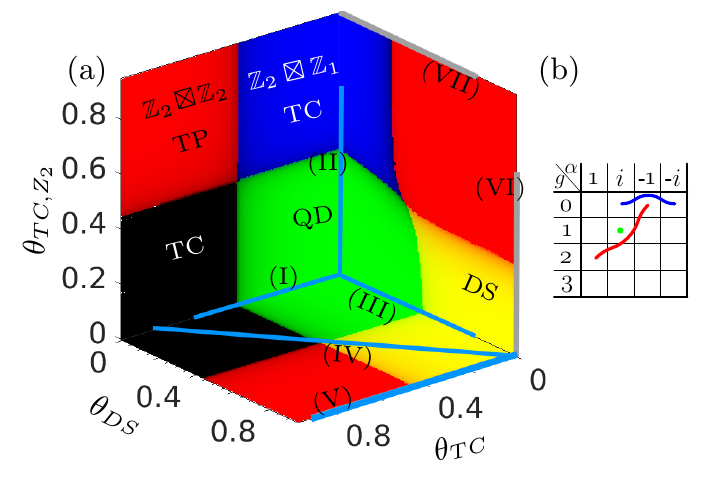}
	\caption{
\textbf{(a)} Phase diagram of the three-parameter family of topological models with
$\mathbb Z_4$ symmetry introduced in Sec.~\ref{subsec:3par-family-def}
which exibits all topological phases, shown along three hyperplanes. The
color coding uses RGB values given by the anyon wavefunction
norms/overlaps shown in \textbf{(b)}, cf.~Appendix~\ref{app:RGB}.  $\chi=16$ has
been used for the approximation of fixed points in the iMPS calculations.}
\label{fig:phasediagram}
\end{figure}

The family in Eq.~\eqref{eq:z4phasediag} is parametrized by three
parameters $\bm{\theta}=\left(\theta_{\text{TC}}, \theta_{\text{TC},Z_2},
\theta_{\text{DS}} \right)$. Fig.~\ref{fig:phasediagram} shows a section
through the phase diagram along the three hyperplanes on which any one of
the three $\theta_\bullet=0$.  The family includes the \znqd4 phase
(green), two toric code phases -- a \zz{4}{2} \tc\ (black) and a \zz{2}{1}
\tc\ (blue), a \ds\ model with \zz{4}{2} symmetry (yellow), and a
\zz{2}{2} trivial phase (red). The color is based on RGB values given by
the order parameters $\langle2,1|0,-1\rangle$ (red),
$\langle1,i|1,i\rangle$ (green), and $\langle0,i|0,-i\rangle\}$ (blue),
which allow to distinguish all those phases, cf.~Appendix~\ref{app:RGB}.

Before we discuss the individual phase transitions in detail, let us give
an overview of our findings:

The majority of the transitions in the phase diagram are governed by the
breaking of a single $\mathbb Z_2$ symmetry in the transfer operator,
corresponding to the arrows in Fig.~\ref{fig:diagLayout}.
Moreover, except for the DS model the fixed points on both sides of the
transition do not exhibit non-trivial SPT order.  Specifically, this
encomasses the \znqd4 QD (with symmetry \zz41) to TC transition (for both
\zz42 and \zz21 TC), as well as the transitions from both TC models to the
trivial phases.  For all of these transitions, we find that they fall
in the 2D Ising universality class, in accordance with the fact that they
are described by a one-dimensional transfer operator undergoing a $\mathbb
Z_2$ symmetry breaking transition. This includes in particular the
transitions marked (I), (II), and (V) in Fig.~\ref{fig:phasediagram}.
This behavior is robust also when considering direct interpolations of the
transfer operator, rather than the interpolation shown in
Fig.~\ref{fig:phasediagram}.

The transitions involving the DS model, on the other side, are more rich.
On the one hand, there are transitions which are again described by the
breaking of a single $\mathbb Z_2$ symmetry, namely the \znqd4 QD (\zz41)
to DS (\zz42) transition, as well as the DS to trivial transition.
Different from the previous case, however, the fixed point at the DS side
of the phase transition exhibits non-trivial SPT order.  This is reflected
in the universality class of the transitions: While the \znqd4 QD to DS
transition along line (III) is still Ising-type, the transition from the
DS (\zz42) to the \zz44 trivial phase (not part of
Fig.~\ref{fig:phasediagram}) seems to belong to the 4-state Potts
universality class. Finally, the DS to \zz22 trivial phase transition in
Fig.~\ref{fig:phasediagram} exhibits continuously varying critical
exponents when moving between the lines (V) and (VI) along the
$\theta_{DS}=1$ plane, whose behavior does not seem to match known
universality classes.

Finally, there are transitions between the DS model and TC models.  There
are two different types: First, the \zz42 DS to \zz21 TC transition, which
corresponds to the breaking of \emph{two} $\mathbb Z_2$ symmetries.
While Fig.~\ref{fig:phasediagram} does not exhibit such a transition, it
is possible to obtain it by linear interpolation of the transfer operator.
The transition is second order and lies in the universality class of the
4-state Potts model, in accordance with the breaking of a $\mathbb
Z_2\times \mathbb Z_2$ symmetry.  On the other hand, there is the
transition between the \zz42 DS and the \zz42 TC, which does not involve
any symmetry breaking, but rather a re-ordering of the fixed point from a
trivial to an SPT phase.  As we have seen, this transition, when realized
by direct interpolation, is 1st order; however, one can also realize a
fine-tuned 2nd order transition through the quadro-critical point in the
bottom plane of Fig.~\ref{fig:phasediagram}, line (IV), in which case a 2D
Ising transition is observed.

Let us now discuss of findings for the individual transitions in detail.

\subsection{Transitions between \znqd{4} QD and $\mathbb Z_2$ topological
phases
\label{subsec:results_z4_z2}}

\subsubsection{\label{sec:num:qd-to-42tc}%
Transition between \znqd{4} \qd\ and \zz{4}{2} \tc}
\begin{figure}[t]
	\includegraphics{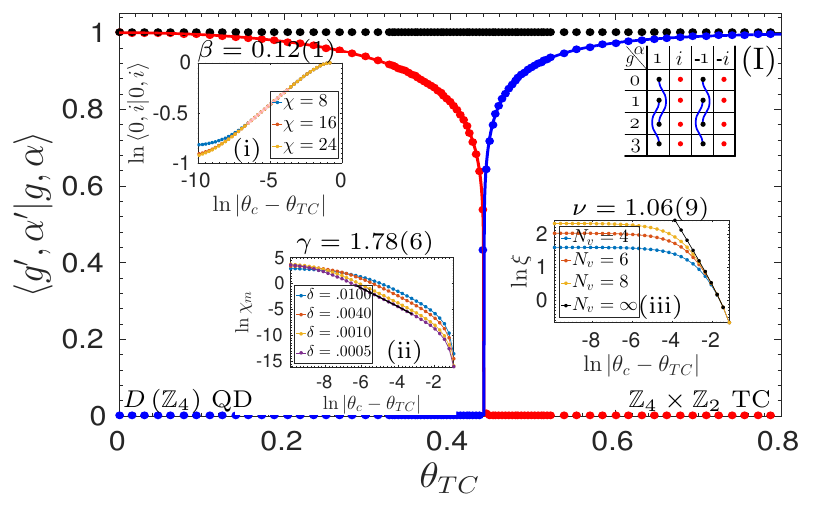}
	\caption{
Condensate and deconfinement fractions for the phase transition between
\znqd{4} \qd\ and \zz{4}{2} \tc\ constructed by local filtering [line (I)
in Fig.~\ref{fig:phasediagram}, computed with $\chi=24$. The color coding
is given by the table in the top right corner as explained in 
Fig.~\ref{fig:compare_v1}. 
(i) Scaling of the deconfinement fraction $\langle 0,i|\langle 0,i\rangle$
in the vicinity of the transition, we find a critical exponent
$\beta=0.12(1)$.  (ii) Critical exponent $\gamma=0.178(6)$ obtained from
the scaling of the susceptibility of the deconfinement, where $\delta$
denotes the step sizes used for approximating of derivative.  
(iii) Correlation length $\xi$ determined from finite cylinders, 
yielding a critical exponent $\nu=1.06(9)$. See text and
Fig.~\ref{fig:compare_v1} for further discussion of methods.
}
\label{fig:QD_TC_Ising}
\end{figure}

A transition between \znqd{4} \qd\ and \zz{4}{2} \tc\ can be obtained
either by local filtering or by direct interpolation of the transfer
operator, as described in Eq.~\eqref{eq:Z4_Z2xZ4TCPT} and
Eq.~\eqref{eq:LinearInt3}, respectively.  We have already considered the
transition obtained by direct
interpolation when introducing our methods in Sec.~\ref{sec:num-methods},
where we found a second-order transition in the 2D Ising universality
class, Fig.~\ref{fig:compare_v1}b.  In the following, we discuss the phase
transition obtained by local filtering along the path labeled by (I) in
Fig.~\ref{fig:phasediagram}a. 

An important feature of this interpolation is that we can devise a
microscopic mapping to the 2D Ising model, including an explicit mapping
of condensate and deconfiment fractions to order parameters and twisted
boundary conditions in the Ising model, see
Appendix~\ref{app:IsingMapping}; it thus allows us to benchmark our
numerical methods with respect to the analytical results.

Fig.~\ref{fig:QD_TC_Ising} gives the condensate fractions as indicated in
the legend on the top right, identical to those in Fig.~\ref{fig:compare_v1}b, as
a function of the interpolation variable $\theta_{\text{\tc}}$.  At the
phase transition, the anyons $|*,\pm i\rangle$ become confined (red),
while $|2,1\rangle$ condenses (blue). The behavior of the order parameters
clearly indicates a second-order phase transition.  The numerical data
(dots) and analytical data (lines) show excellent agreement, and the
critical point is in agreement with the analytical value
$\theta_c=\frac{1}{2}{\ln\left(1+\sqrt{2}\right)}$.

From the order parameters, we can extract the critical exponent $\beta
=0.12(1)$, consistent with the analytical value $1/8$,
Fig.~\ref{fig:QD_TC_Ising}(i). A further critical exponent can be
obtained by studying the suceptibility of the order parameter to an
external ``field'', cf.~Sec.~\ref{subsec:methods_susceptibility}: Here,
the order parameter measures the spontaneous breaking of the
$(\openone,X^2)$ symmetry.  It can be
explicitly broken by
modifying the filtering tensor (\ref{eq:Z4_Z2xZ4TCPT}) as
\begin{equation}\label{eq:Z4_Z2xZ4TCPT_h}
\cbox{3}{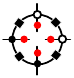}\ ,
\end{equation}
where 
$\cbox{3.5}{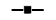}:=\mathrm{diag}(1+h,1-h,1-h,1+h)$.
The susceptibility is then defined as 
\begin{equation} 
\chi_m(\theta) := \left.
\frac{\partial {\left \langle 0,i | 0,i\right\rangle} }{\partial h}
\right|_{h=0} \ ,
\end{equation}
where $|0,i\rangle$ is a function of $\theta$ and $h$.  We have examined
the behavior of $\chi_m$ by using finite differences for the derivative
for different step sizes $\delta$. The scaling of $\chi_m$ with respect to
$\theta$ close to the critical point, Fig.~\ref{fig:QD_TC_Ising}(ii), 
gives  a critical exponent $\gamma=1.78(6)$, consistent with the
analytical value $7/4$. Finally, we have also determined the correlation
length on an infinite cylinder, yielding a critical exponent
$\nu=1.06(9)$, Fig.~\ref{fig:QD_TC_Ising}(iii), in agreement with the
analytical value $\nu=1$.

\subsubsection{Transition between \znqd{4} \qd\ and \zz{2}{1} \tc}

\begin{figure}[!t]
	\includegraphics{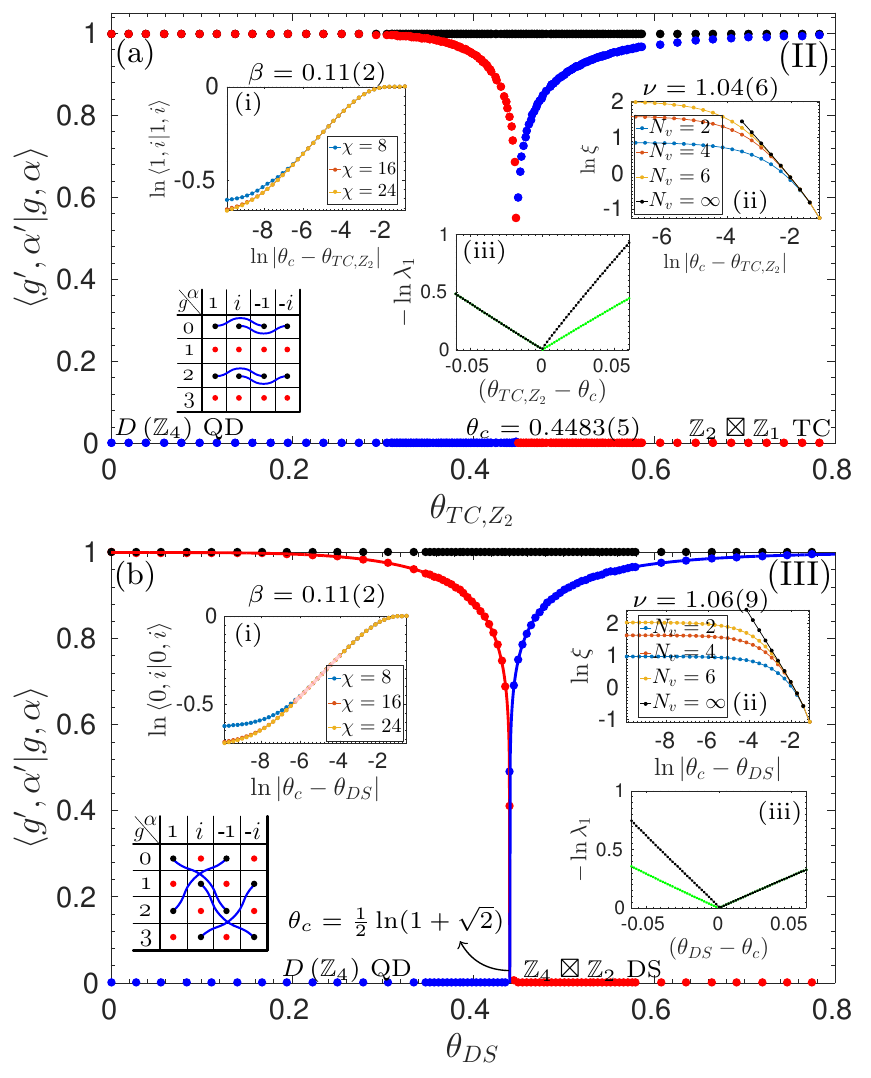}
	\caption{
Order parameters for the phase transitions between \textbf{(a)} \znqd{4}
\qd\ and \zz{2}{1} \tc\ and \textbf{(b)} \znqd{4} \qd\ and \zz{4}{2} \ds.
The colors are defined through the anyon tables in the lower left corners,
cf.~Fig.~\ref{fig:compare_v1}. Insets (i) and (ii) in (a) and (b)
give the extraction of the critical exponents $\beta$ and $\nu$ of an
order parameter and correlation length, respectively. Inset (iii) in (a)
and (b) shows the inverse correlation length extracted from the transfer operator
using an iMPS excitation ansatz.  Here, black (green) lines correspond to
correlations of anyons with trivial (non-trival) flux. Cf.~text for a
discussion.} \label{fig:QD_DSTC}
\end{figure}

Let us now consider the phase transition between \znqd{4} \qd\ and
\zz{2}{1} \tc\ labeled (II) in Fig.~\ref{fig:phasediagram}a; 
the \zz{2}{1} \tc\ is obtained from \znqd4 \qd\ by condensing the
$|0,-1\rangle$ charge rather than the $|2,1\rangle$ flux 
as for the \zz42\ TC, leading to the confinement of the fluxes 
$\lbrace |1,*\rangle, |3,*\rangle \rbrace$. 

Fig.~\ref{fig:QD_DSTC}a summarizes the numerical results on the
interpolation.  The main panel shows the condensation and deconfinement
fractions, as indicated in the bottom left. The data is consistent with a
2nd order phase transition at $\theta_c = 0.4483(5)$. The scaling of the
order parameters $|\langle 1,i|1,i\rangle|$  yields a critical exponent
$\beta=0.11(2)$ [inset (i)], and for the correlation length, we find
$\nu=1.04(6)$ [inset (ii), from cylinders], both consistent with the 2D
Ising universality class. Inset (iii) shows the inverse correlation length as
extracted from the transfer operator using an excitation ansatz.  Here,
green dots correspond to topological excitations with a flux string
attached (i.e., domain wall excitations of the broken symmetry
\zz41$\to$\zz21), and black dots to zero-flux excitations (both with and
without charge); we thus find
that the dominating length scale after the transition indeed arises from
the confinement length of a flux (or dyon). 

Let us add that we found that the phase transition between \znqd{4} \qd\
and \zz{2}{1} \tc\ constructed through direct interpolation of the
transfer operator to lie in
the Ising universality class as well.

\subsubsection{Transition between \znqd{4} \qd\ and \zz{4}{2} \ds}

As a last transition out of the \znqd4\ QD phase, we consider the
transition to the \zz{4}{2} \ds\ model via the path (III) in Fig.
\ref{fig:phasediagram}a, Eq.~\eqref{eq:Z4_Z2xZ4DSPT}. This transition can
yet again be mapped to the 2D Ising model,
cf.~Appendix~\ref{app:IsingMapping}. The results are shown in Fig.
\ref{fig:QD_DSTC}b, were in the main panel dots (lines) give the numerical
(analytical) result: Numerical and analytical order parameters show
excellent agreement, and we find a second order phase transition whose
critical exponents match those of the 2D Ising model, with the transition
at $\theta_c=\frac{1}{2}{\ln\left(1+\sqrt{2}\right)}$.  In particular,
inset (iii) shows again the subleading eigenvalue of the transfer
operator, where green dots label sectors with a non-trival flux string;
the dominant length scale before the transition thus arises from the mass
gap of the $|2,-1\rangle$ dyon which is condensed in the DS phase.

\subsection{Phase transitions between toric codes and double semion model
\label{subsec:results_tc_ds}}

Let us now turn towards phase transitions between the Toric Code and the
Double Semion phase.  This transition is of particlar interest, as it
is not described by anyon condensation, and it has been conjectured that
it should thus be first order, which is supported by exact diagonalization
calculations~\cite{morampudi2014numerical}.

\subsubsection{Transition between \zz{4}{2} \tc\ and \zz{4}{2} \ds }

Unlike for phase transitions which are described by condensation of
anyons, obtaining an interpolation which achieves a direct transition
between the TC and the DS phase is non-trivial and requires fine-tuning --
for a generic interpolation, one would expect to go through an
intermediate phase which has condensation-driven transitions to either TC
and DS, i.e.\ either a trivial or a \znqd4\ phase.

We had already earlier studied one direct transition between \zz{4}{2}
\ds\ and \zz{4}{2} \tc\ in Sec.~\ref{sec:num-methods},
constructed by direct interpolation of the transfer operator, where we
found that the transition was first order, cf.~Fig.~\ref{fig:compare_v1}a
and Fig.~\ref{fig:compare_corrs}a,c,e.

\begin{figure}[t]
	\includegraphics[width=26em]{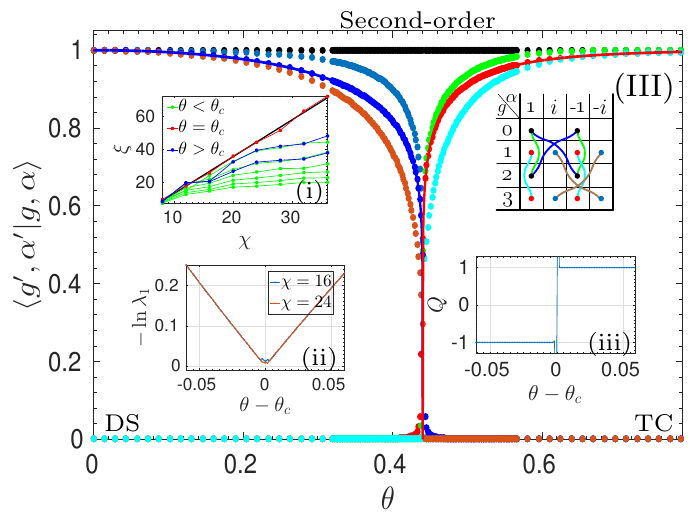}
	\caption{
Order parameters for the phase transition between \ds\ and \zz42 \tc\
along the line (IV) in Fig.~\ref{fig:phasediagram}a, computed with
$\chi=24$. The colors are defined through the table on the right,
cf.~Fig.~\ref{fig:compare_v1}.
(i) Scaling of the correlation length $\xi$ of the boundary phase,
indicative of a second-order transition.. (ii) Correlation length
determined from the transfer operator using an iMPS excitation ansatz.  
(iii) Order parameter for SPT order in the fixed point of the transfer
operator, Eq.~\eqref{eq:spt-order-par}, in the vicinity of the phase
transition.  It exhibits a sharp jump which allows to accurately determine
the transition point.} \label{fig:Z2xZ4TC_DS_Ising}
\end{figure}

Another possibility of obtaining a direct transition is to consider the
horizontal plane ($\theta_{\text{TC},Z_2}=0$) in the phase diagram
Fig.~\ref{fig:phasediagram}, which exhibits a quadro-critical point in
which TC, DS, trivial, and \znqd4\ phases meet.  As mentioned earlier, the
whole plane can be mapped to the 2D Ising model, and so can a diagonal
path $\bm \theta(\theta) = \left(\theta,0,1-(1-\theta_c)\theta/\theta_c
\right)$ [with $\theta_c = \frac{1}{2}\ln(1+\sqrt{2})$], labeled (IV) in
Fig.~\ref{fig:phasediagram}a, which passes through the critical point at
$\bm\theta(\theta_c) = (\theta_c,0,\theta_c)$. The numerical findings
along this interpolation are shown in Fig.~\ref{fig:Z2xZ4TC_DS_Ising}a and
are consistent with a phase transition in the 2D Ising universality class.

While the boundary states $|\ell)$, $|r)$ of the \tc\ and \ds\ model both
have \zz{4}{2} symmetry, they differ in the projective action $V_{\bm g}$,
$V_{\bm h}$ of the generators $\bm g=(X,X)$ and $\bm
h=\left(\mathds{1},X^2\right)$ on the virtual indices of boundary MPS,
cf.~Eqs.~\eqref{eq:z4xz2_symmprop} and \eqref{eq:ds_symmprop}: While in
the case of the TC phase, the symmetry actions commute, in the DS phase
they form a non-trivial projective representation equivalent to the Pauli
matrices. This is in close analogy to the trivial vs.\ Haldane phase in
the case of $\mathbb Z_2\times \mathbb Z_2\subset \mathrm{SO}(3)$ symmetry
for 1D spin chains.  These two phases can be distinguished by an order
parameter $\mathrm{tr}[V_{\bm g} V_{\bm h} V_{\bm g}^\dagger V_{\bm
h}^\dagger]=\pm 1$ which measures the commutator of the virtual symmetry
actions. It can be computed from the iMPS description of $|\ell)$ by
considering the normalized fixed points $\sigma_{g,g'}$ of its dressed channel
operators $\mathbb F_{g,g'}$ (see Fig.~\ref{fig:num_1}b) as
\begin{equation}
\label{eq:spt-order-par}
Q = \text{Tr}\left( \sigma_{1,1}\sigma_{0,2} {\sigma^{-1}_{1,1}} {\sigma^{-1}_{0,2}} \right)\ ,
\end{equation}
given that the iMPS is in canonical form with
$\sigma_{0,0}\propto\openone$ (then, $\sigma_{g,g'}\propto V_{g,g'}$ with
$V_{g,g'}$ unitary, cf.\ the discussion in Sec.~\ref{subsec:order-params}).
Here, a value of $Q=+1$ ($Q=-1$) indicates that the system is in the TC
(DS) phase~\cite{haegeman2012order,pollmann:spt-detection-1d}.
Fig.~\ref{fig:Z2xZ4TC_DS_Ising}a(iii) shows $Q$ in the vicinity of phase
transition: It exhibits a sharp jump, which allows to accurately determine
the value of the critical point.

\subsubsection{Transition between \zz{2}{1} \tc\ and \zz{4}{2} \ds%
\label{subsec:tc21-to-ds}}

\begin{figure}[t]
	\includegraphics[width=26em]{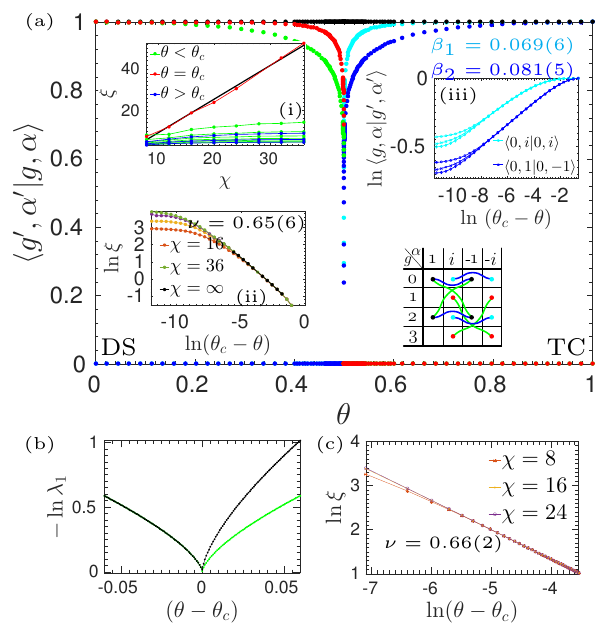}
	\caption{
\textbf{(a)} Order parameters along the phase transition between \zz{4}{2}
\ds\ and \zz{2}{1} \tc, cf.~Fig.~\ref{fig:compare_v1} for the legend. The
model has an exact self-duality $\theta\leftrightarrow1-\theta$, see text.
(i) Correlation length (from iMPS) vs.\ $\chi$ around the transition,
indicative of a 2nd order transition. (ii) extraction of critical exponent
$\nu$ from finite cylinders, yielding $\nu=0.65(6)$. 
(iii)~Extraction of critical exponents $\beta$, yielding two exponents
$\beta_1=0.069(6)$ and $\beta_2=0.081(5)$. The observed exponents are
compatible with a $4$-state Potts transition. \textbf{(b)} Inverse
correlation length extracted from the transfer matrix using an iMPS
excitation ansatz. Black (green) denotes again anyon correlations with
(without) flux, showing that the self-duality map exchanges flux
and charge.
\textbf{(c)} Correlation length computed using iMPS, yielding
$\nu=0.66(2)$.  
} \label{fig:DS_Z2TC}
\end{figure}

In contrast to the previous case, there does not exist a direct path
between \zz{2}{1} \tc\ and \zz{4}{2} \ds\ in Fig. \ref{fig:phasediagram}a
on the $\theta_{\text{TC}}=0$ hyperplane.  We can however obtain a direct
phase transition between the two phases by linear interpolation of the
on-site transfer operators, cf.~Eq.~\eqref{eq:LinearInt3}.  The results
are shown in Fig.~\ref{fig:DS_Z2TC}: We find clear
signs of a second order phase transition from DS to TC driven by
simultaneous condensation of the $|0,-1\rangle$ anyon and de-condensation
of the $|2,-1\rangle$ anyon, which is witnessed by a diverging correlation
length and continuously vanishing order parameters.  

The critical point is found at $\theta_c=0.5$, which we can trace back to
a self-duality of the model.  Specifically, there exists a Matrix Product
Unitary (MPU) $U$ which interchanges the on-site transfer operator of the
DS and the TC fixed point when commuted with it; this implies that for the
transfer operator $\mathbb T(\theta)$ of a column, $U \mathbb{T}(\theta)
U^\dag=\mathbb{T}(1-\theta)$.  The explicit construction and analysis of
the MPU $U$ is given in Appendix \ref{app:z2-dsmapping}. In fact, $U$ also
interchanges the order parameters for the two phases, and thus, the order
parameters in Fig.~\ref{fig:DS_Z2TC} are fully symmetric.

From the scaling of the correlation length at the critical point we
extract a critical exponent $\nu\approx 0.66$. The order parameters
exhibit two different critical exponents, which we determine as 
$\beta_1 = 0.069(6)$ (for the deconfinement fraction $\langle
0,i|0,i\rangle$) and $\beta_2 = 0.081(5)$ (for the 
condensate fraction $\langle 0,-1| 0,1\rangle$), respectively. Our
findings for $\nu$ and $\beta_2$ are in accordance with the universality class
of the Ashkin-Teller model at the $4$-state Potts point (with $\nu=2/3$
and $\beta=1/12$), which is in agreement with the simultaneous breaking of
two $\mathbb Z_2$ symmetries at the transition.

\subsection{Phase transition with continuously varying critical exponents
\label{subsec:results_contvar}}

\begin{figure}[t]
	\includegraphics[width=22em]{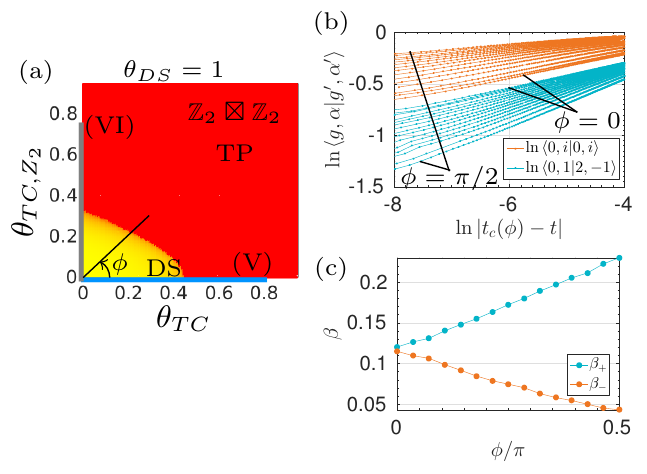}
	\caption{
\textbf{(a)} $\theta_{\text{DS}}=1$ hyperplane of the three-parameter
family constructed in Sec.~\ref{subsec:3par-family-def},
Fig.~\ref{fig:phasediagram}, exhibiting a DS and a \zz22 trivial phase.
Cf.~Fig.~\ref{fig:phasediagram} for the color coding. 
Transitions are scanned along
lines with different angles $\phi$.  \textbf{(b)} Scaling of $\langle
0,i|0,i\rangle$ and $\langle 0,1|2,-1\rangle$ in the vicinity of the
transition, as a function of $\phi$. \textbf{(c)} Critical exponents
$\beta_{\pm}$ for their scaling as a function of $\phi$; we find a
continuously varying transition with $\nu\approx 1$ constant.
}
\label{fig:KT_plane}
\end{figure}

An interesting feature of the phase diagram of Fig.~\ref{fig:phasediagram}
is the transition between the DS and the trivial phase in the
$\theta_{DS}=1$ hyperplane spanned by the lines (V) and
(VI) in Fig.~\ref{fig:phasediagram}. A cut through this hyperplane is
shown in Fig.~\ref{fig:KT_plane}a. When moving along the plane, as
parametrized by the angle $\phi$, $\left(\theta_{\text{TC}},
\theta_{\text{TC},Z_2}\right) = t\left(\cos\phi,\sin\phi \right)$, we find
that the transition is second order with critical exponent $\nu=1$, but
the critical expnents $\beta_{\pm}$ for the order parameters on the two
sides of the transition change continuously.  This is shown in
Fig.~\ref{fig:KT_plane}b,c.  Here, $\beta_+$ is the critical exponent of
the order parameter $\langle 0,1|2,-1\rangle$ in the DS phase, and 
$\beta_-$ is the critical exponent of the order parameter $\langle
0,i|0,i\rangle$ in the trivial phase. At $\phi=0$, the transition is in
the Ising universality class with $\beta_{+}(\phi=0)=\beta_-(\phi=0)=1/8$.
As we change $\phi$, $\beta_+$ grows until the final value
$\beta_+(\phi=\pi/2)=0.23(1)$, while $\beta_-$ decreases until
$\beta_-(\phi)=0.04(1).$ Let us add that the critical behavior is
independent of the direction along which one crosses the phase transition,
as to be expected.  

Given the symmetries of the model, it is plausible to conjecture that this
transition maps to the self-dual line of the Ashkin-Teller (AT) model
which exhibits continuously varying critical exponents as well, including two
different ``electric'' and ``magnetic'' exponents $\beta_e$ and
$\beta_m$~\cite{baxter:book}.  However, there are several discrepancies,
such as the constant $\nu=1$ as opposed to a continuously varying $\nu$
in the AT model, and the fact that in the AT model, $\beta_e$ and
$\beta_m$ both change in the same direction, whereas $\beta_+$ and $\beta_-$
change in opposite directions, leaving the identification of
the exact nature of this transition an open question.

\subsection{Phase diagrams of toric codes and double semion model}\label{subsec:PhDiag_TCDS_TPs}

\begin{figure}[b]
	\includegraphics[width=25em]{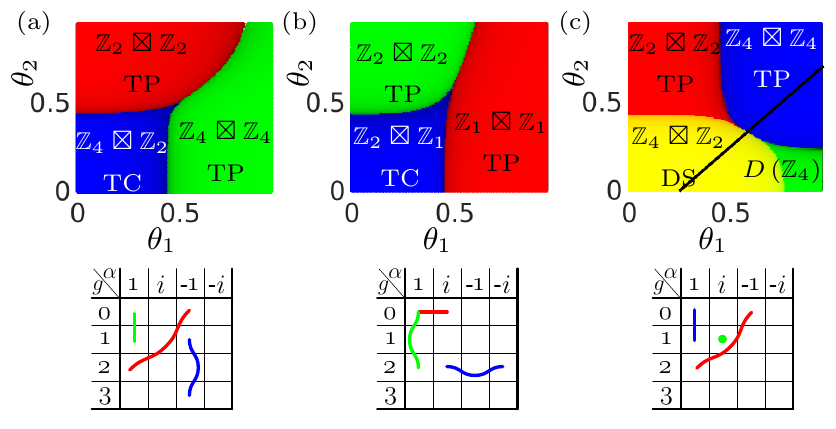}
	\caption{
Phase diagrams of models which are obtained by 
deforming \textbf{(a)} \zz{4}{2} \tc,  \textbf{(b)} \zz{2}{1} \tc, and
\textbf{(c)} \zz{4}{2} DS towards trivial phases, as discussed in 
Sec.~\ref{subsec:interpolations_introtrivial}.
The corresponding anyon tables below each phase diagram explain the color
coding, where the RGB values of each point are given by the corresponding
anyon wavefunction overlaps/norms. 
All data has been obtained with $\chi=16$.}
 \label{fig:TP_Ph}
\end{figure}

After having studied the phase diagram of the three-parameter family in
detail, we will now proceed to examine the behavior of phase transitions
which can have been constructed in
Sec.~\ref{subsec:interpolations_introtrivial} by further deforming the
toric codes and double semion model down to trivial phases.

\subsubsection{\zz{4}{2} Toric Code}

In the case of \zz{4}{2} toric code, the two-parameter deformation
Eq.~\eqref{eq:Z2xZ4TCTP_PT} can induce phase transition to either the
\zz{4}{4} \tp\ or \zz{2}{2} \tp.  The phase diagram of the model is shown
in Fig.~\ref{fig:TP_Ph}a).  Away from the tri-critical regime (where
convergence becomes slow), we find that the phase transitions between the
\zz{4}{2} \tc\ and either of the trivial phases lies in the Ising
universality class.

\subsubsection{\zz{2}{1} Toric Code}

The two-parameter family of Eq.~\eqref{eq:Z2TCTP_PT} can drive the
\zz{2}{1} \tc\ into the two trivial phases with \zz{1}{1} and  \zz{2}{2}
symmetry, respectively.  The phase diagram is shown in
Fig.~\ref{fig:TP_Ph}b. At $\theta_1=0$, the phase transition between
\zz{2}{1} \tc\ and \zz{2}{2} \tp\ can be mapped to the 2D Ising model.
Furthermore, away from the tri-critical regime, the transitions between
\zz{2}{1} \tc\ and the two trivial phases are found to lie in the Ising
universality class.

\subsubsection{\zz{4}{2} Double semion model}\label{subsubsec:dsmodel}

\begin{figure}[b]
\includegraphics[width=0.9\columnwidth]{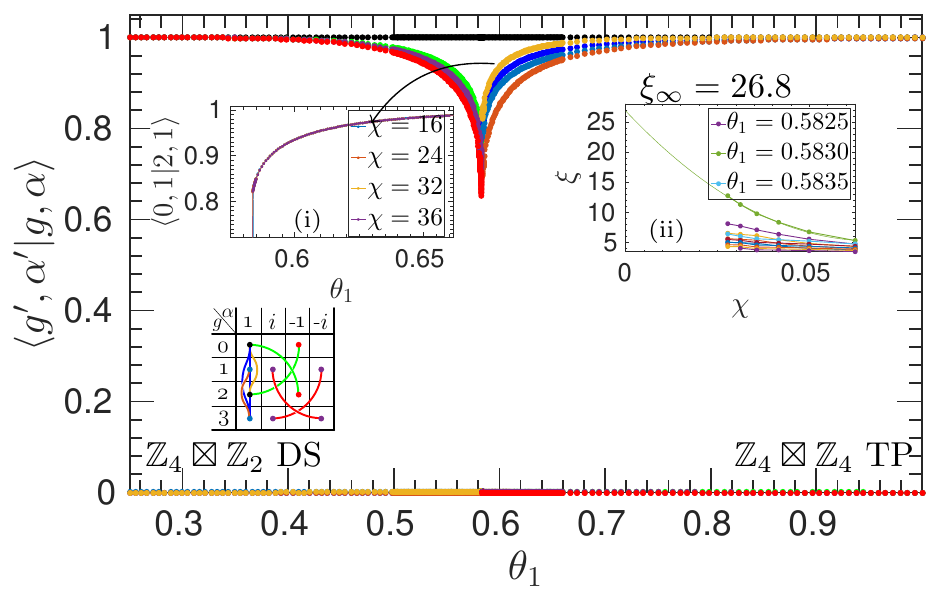}
\caption{Order parameters along the interpolation DS to \zz44 trivial
phase (cf.~table in bottom left). The left inset shows a zoom of the
condensate fraction $\langle 0,1|2,1\rangle$ in the vicinity of the
transition, and the right inset the
scaling of the correlation length with the iMPS bond dimension $\chi$
(with a quadratic fit), both of which show clear signs of a first order
transition.} \label{fig:ds_z4z4}
\end{figure}

Let us now turn to the two-parameter family of Eq.~\eqref{eq:DSTP_PT}. It
exhibits a \zz42 DS phase, two trivial phases (\zz44 and \zz22), as well
as the \znqd4 QD phase. Fig.~\ref{fig:TP_Ph}c shows the phase diagram. While
the  transitions between \zz{4}{2} \ds\ and \znqd{4} \qd\ across the
horizontal axis and between \zz{4}{2} \ds\ and \zz{2}{2} \tp\ across the
vertical axis lie in the Ising universality class, the transition from
\zz42 DS to the \zz44 trivial phase is different: One the one hand, it
requires fine-tuning to achieve a direct transition, which we obtain along
the black line in Fig.~\ref{fig:TP_Ph}c
given by $\theta_2 = \theta_1 - (\theta_1^T-\theta_2^T)$
through the transition point $(\theta_1^T, \theta_2^T)=(0.5830,0.3313)$,
and on the other hand, it exhibits clear signs of a first order transition,
as shown in Fig.~\ref{fig:ds_z4z4}.

\section{Conclusions}\label{sec:conclusions}

In this paper, we have used the framework of PEPS to study topological phase
transitions.  Using the formalism of $G$-injective
PEPS, we have to set up families of models which interpolate between different
topological phases, and have utilized the description of
topological excitations in PEPS through string operators on the entanglement
degrees of freedom to set up order parameters 
characterizing condensation and deconfinement of anyons, which allowed us to
study the topological phases of these models and the transitions between
them.

Starting from a model with $\mathbb Z_4$ symmetry, we have obtained a family
of states encompassing the \znqd4 quantum double, toric code, double
semion, and trivial phases, and set up interpolations between them.
Using order parameters for condensation and deconfinement, anyonic
correlation functions, and some further probes, we have characterized the
phase diagram of the model. We found a rich structure where all possible
phases and the transitions between them are realized.  Analyzing the
phase transitions revealed a range of different types of transitions, both
first and second order. We
found a number of transitions in the 2D Ising universality class,
compatible with the understanding that these transition break a single
$\mathbb Z_2$ symmetry, but also transitions in the 4-state Potts
universality class, as well as transitions with continuously varying
exponents whose universal behavior is not yet identified. We also found
that the transition between double semion and toric code could be both
first and second order, and exhibit different critical exponents.

It would be interesting to further investigate the nature of a generic
interpolation between the toric code and the double semion model.  If one
would find that an interpolation between these phases (or any other two
phases) is generically first order, this would also have implications on
the results obtained with fully variational PEPS calculations, where the
tensor could change abruptly at the phase transition: Having generically a
first order transition implies that there is a range of values of the 
order parameters which cannot be reached by any choice of parameters,
suggesting that the first order transition will persist even if considering
a fully variational simulation.

The order parameters employed in this work for the analysis of topological
phase transitions are not restricted to explicitly designed families of
tensors: They can also be applied to scenarios where the tensors are
obtained variationally by minimizing the energy of a given Hamiltonian, as
long as the numerical method keeps track of the different topological
symmetry sectors in the tensor. Note that this does not rule out explicit
breaking of the symmetry which is important to obtain the best variational
wavefunctions, as long as the sector label of the symmetry broken tensor
is being kept track of as well. It would thus be interesting to use these
order parameters for condensation and confinement to analyze the behavior
of further topological phase transitions through variational PEPS
calculations.

\begin{acknowledgments}
MI thanks Manuel Rispler for helpful discussions.  
Part of the computations were performed on the JARA-HPC cluster at RWTH
Aachen University, supported by JARA-HPC grant jara0092. This work was
supported by the DFG through Graduiertenkolleg 1995, and the European
Union through the ERC Starting Grant \mbox{WASCOSYS} (No.~636201).
\end{acknowledgments}

\onecolumngrid

\vspace*{1cm}
\hspace*{\fill}\rule{15cm}{0.5pt}\hspace*{\fill}
\vspace*{1cm}

\twocolumngrid

\appendix

\section{Phase transitions and fidelity per site} \label{app:fid}
\begin{figure}[!t]
	\includegraphics[width=26em]{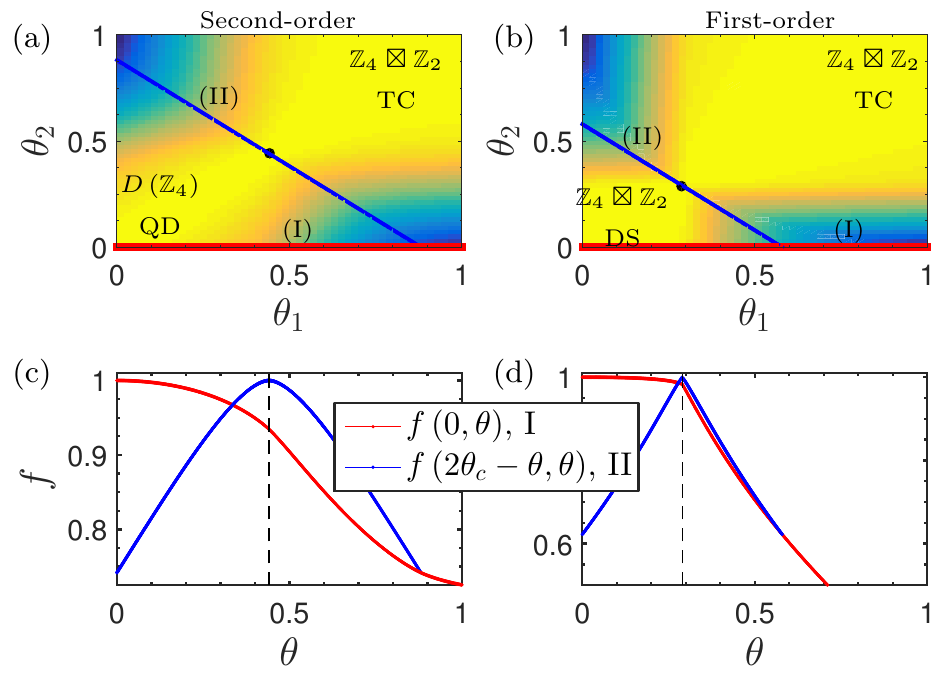}
	\caption{Comparison of fidelity per site for the second order phase transition by \znqd{4} \qd\ and \zz{4}{2} \tc\ (a,c) and first order phase transition between \zz{4}{2} \ds\ and \zz{4}{2} \tc\ (b,d).}
	\label{fig:fidelity_per_site}
\end{figure}

The notion of fidelity per site has been studied in the context of tensor network states in \cite{fid4}. It gives a measure of distinguishably between quantum states and it is defined as the normalized overlap of two wavefunctions per site.
\begin{equation}
\begin{aligned}
f(\theta_1,\theta_2)&={\left|{\frac{\langle \theta_2 | \theta_1 \rangle}{\sqrt{\langle \theta_1 | \theta_1
				\rangle\langle \theta_2 | \theta_2 \rangle}}}\right|}^{1/N},
\end{aligned}
\end{equation}
and ${N\rightarrow \infty}$ in the thermodynamic limit. Let
\begin{equation}\label{eq:fid_H}
\mathbb{H}\left(\theta_1,\theta_2\right) := \cbox{4.5}{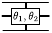}\ ,
\end{equation}
where
\begin{equation}
\cbox{10}{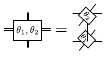}
\end{equation}
is the on-site transfer operator and the top and bottom tensors in Eq. \eqref{eq:fid_H} are the local tensors for the left and right fixed point of the transfer operator $\mathbb{T}(\theta_1,\theta_2)$ computed using iMPS algorithm. Let $\lambda_0$ be the largest eigenvalue of $\mathbb{H}$, then 
\begin{equation}
f(\theta_1,\theta_2) =
\frac{\lambda_0(\theta_1,\theta_2)}{\sqrt{\lambda_0(\theta_1,\theta_1)\lambda_0(\theta_2,\theta_2)}}\ ,\ \text{as} \
N\rightarrow \infty
\end{equation}
Fidelity per site can used to characterize the behavior of phase transitions. Fig. \ref{fig:fidelity_per_site} show a comparison of fidelity per site for (a) second- and (b) first-order phase transition. Fig. \ref{fig:fidelity_per_site}(c,d) shows the behavior of fidelity per site across different slices marked in the surface plot Fig. \ref{fig:fidelity_per_site}(a,b). Although, $f(\theta_1,\theta_2)$ changes smoothly in the first-order case we observe a cusp like behavior in the transition regime which is qualitatively different in comparison to the second order phase transition between the \znqd{4} \qd\ and \zz{4}{2} \tc.

\section{\label{app:dispersion}%
Excitation spectrum of the transfer operator}

We analyze the dispersion relation of the transfer operator. The computation of the low-lying excited states of the transfer operator has been achieved by using the excitation ansatz \cite{haegeman2013post}. We present our findings for the phase transition between \zz{4}{2} \ds\ and \zz{1}{1} \tc\ phase, where the fixed points of the transfer operator spontaneously break the \zz{4}{2} symmetry to \zz{2}{1}. 

Since the transfer operator is \zz{4}{4} invariant, we can label each excitation of the transfer operator for the given $k$ as $\lambda_{g,\alpha}^{g',\alpha'}$, where $(g,g')$ is a label for the conjugacy class and $(\alpha,\alpha')$ is a label for an irrep. of \zz{4}{4}. It is important to note that the different species of anyonic particles in the \tc\ and \ds\ phase can be labeled by $(g,\alpha)$.

\begin{figure}[!t]
	\centering
	\includegraphics[width=26em]{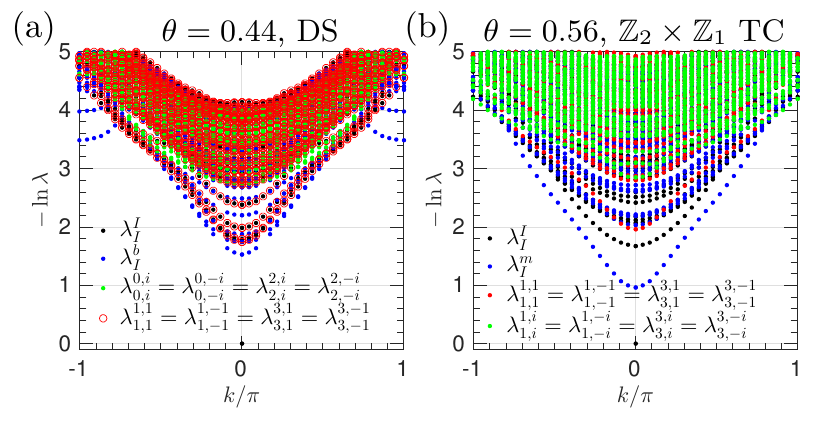}
	\caption{Dispersion relation of the  transfer operator at the data
points in the vicinity of phase transition between \zz{4}{2} \ds\ and
\zz{2}{1} \tc\ phase (see Sec.~\ref{subsec:tc21-to-ds}).
The computations have been
performed for the bond dimension $\chi=24$.} \label{fig:DSZ2TC_Dispersion}
\end{figure}

As the system is tuned from the DS phase towards the critical point, bosonic excitations get condensed to the vacuum, and this property is crucial in determining the behavior of the system. Remarkably as first suggested in \cite{haegeman2015shadows}, that condensation of bosonic anyons is manifested in the excitation spectrum of the transfer operator. The low lying excitations labeled as $\lambda_I^b$ in Fig. \ref{fig:DSZ2TC_Dispersion}a are identified with the condensation of bosons. Furthermore, the excitations labeled as $\lambda_{0,i}^{0,i}$ and $\lambda_{2,i}^{2,i}$ represent the deconfinement of  $e$ and $em$ anyons respectively.

Similarly, on the other side of the critical point in \zz{2}{1} \tc\ phase, the behavior of the system is characterized by the condensation of magnetic anyons (labeled as $\lambda_I^m$ in Fig. \ref{fig:DSZ2TC_Dispersion}b). Excitation labeled as $\lambda_{1,i}^{1,i}$ and $\lambda_{1,-i}^{1,-i}$ manifest the deconfinement of semions and their conjugates.

\section{\label{app:RGB}RGB coding of the phase diagram}

The phases in Fig.~\ref{fig:phasediagram} are encoded using an RGB scheme,
where the values for red, green, and blue are determined by the three
fractions $\mathcal{C}=\{\langle2,1|0,-1\rangle$, $\langle1,i|1,i\rangle,
\langle0,i|0,-i\rangle\}$, whose values are sufficient to visualize every
phase which can be realized by deformation. In the following, we will
explain the appearance of different phases in Fig. \ref{fig:phasediagram}.
\begin{enumerate}
\item In \znqd{4} \qd, none of the possible anyons are condensed or confined, which means that the only fraction from $\mathcal{C}$ with a non-zero value is $\langle1,i|1,i\rangle$. So the green region in the phase diagram is identified with \znqd{4} \qd.

\item In the case of \zz{4}{2} \tc, all the anyons of form $|*,\pm i\rangle$ are confined, which implies that the overlaps $ \langle 1,i|1,i\rangle$ and $\langle 0,i|0,-i\rangle$ are equal to zero. Furthermore, the anyon $|2,1\rangle$ is condensed to the vacuum but it can be distinguished from the anyon $|0,-1\rangle$. Since every fraction in $\mathcal{C}$ is zero for \zz{4}{2} \tc\ phase, every point in the black region corresponds to \zz{4}{2} \tc.

\item In \zz{2}{1} \tc\ phase, anyons of the form $|1,*\rangle$ and $|3,*\rangle$ are confined.
Anyons $|2,1\rangle$ and $|0,-1\rangle$ are not confined but they can be distinguished from each other. The only fraction in $\mathcal{C}$ with a non-zero value is $\langle 0,i|0,-i\rangle$ which explains the blue color for \zz{2}{1} \tc.

\item Anyons $|0,\pm i\rangle$ are confined in \zz{4}{2} \ds\ phase which implies that the overlap $\langle 0,i|0,-i\rangle$ is zero. On the other hand, the anyon $|1,i\rangle$ is deconfined and the anyons $|2,1\rangle$ and $|0,-1\rangle$ are mutually indistinguishable (i.e. $\langle 2,1|0,-1\rangle=1$). The fractions in $\mathcal{C}$ with non-zero value are $\langle 0,i|0,i\rangle$ (green) and $\langle 2,1|0,-1\rangle$ (red). Sum of red and green produces yellow, so the \zz{4}{2} \ds\ phase is identified with yellow region.	

\item In the case of \zz{2}{2} \tp\ the only non-zero fraction from $\mathcal{C}$ with a non-zero value is $\langle 2,1|0,-1\rangle$ which determines the color of \zz{2}{2} \tp\ phase to be red.

\end{enumerate}

\section{Ising model and topological phase transitions}\label{app:IsingMapping}
In this appendix, we discuss a mapping between the partition function of classical Ising model and the norm of vacuum state which is parametrized by the tuning variable $\theta$. We will focus our attention here to the phase transition between \znqd{4} \qd\ and \zz{4}{2} \tc, but the description is generic enough to be applied in other cases.

\subsection{Classical Ising model}
\begin{figure}[t]
	\includegraphics[width=26em]{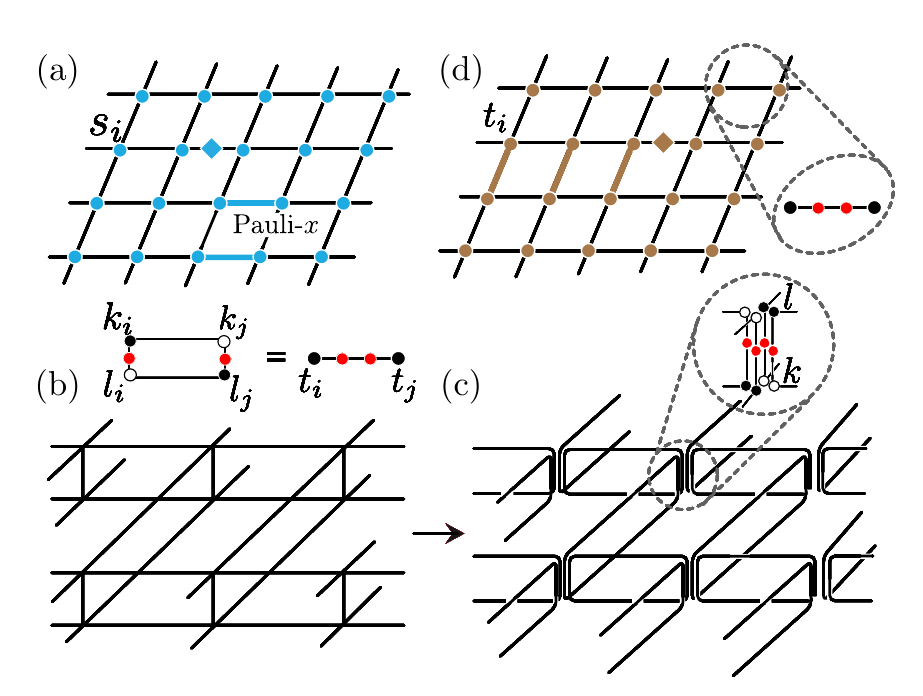}
	\caption{(a) Standard Ising model on a square lattice. The string of defective edges shown by colored edges indicates the presence of Pauli-$x$ in the link. An edge with the diamond denotes an action corresponding to local order parameter. (b) The norm of the vacuum $|0,1\rangle$ constructed by contracting the bra and ket index. (c) More descriptive illustration for the tensor network of vacuum with the local structure of on-site tensors. (d) Ising model which emerged from the norm of vacuum.}
	\label{fig:correspond1}
\end{figure}

We begin by writing down the partition function in terms of Ising variables $s_i$ assigned to each vertex (Fig. \ref{fig:correspond1}a).
\begin{equation}\label{eq:Z}
\mathcal{Z}=\sum _{ \bm{s} }{ \prod _{ \left<i,j \right> }{ { e }^{ \beta { s }_{ i }{
			s }_{ j } } }  }
\end{equation}
For later purposes, it will be convenient to interchangeably use binary variables $b_i=\{0,1\} $ and $s_i=\{-1,1\}$, where $s_i=\left(-1 \right)^{b_i}$, to express each Ising configuration. We use the following graphical notation to represent Boltzmann weights on the horizontal and vertical edges of square lattice.

\begin{equation} \label{eq:isingboltzmanwt1}
\cbox{3}{figures/line_1h} \ \text{or} \cbox{1.5}{figures/line_1v}=
\begin{cases} 
e^{\beta} & \text{if\ }  b_i=b_j \\
e^{-\beta} & \text{otherwise}
\end{cases}
\end{equation}
It is possible to construct a defective edge by inserting a Pauli-$x$ between connecting sites, which modify the Boltzmann weights as follows 
\begin{equation}\label{eq:defectsIsing}
\cbox{3}{figures/line_1e} =
\begin{cases}
e^{-\beta} & \text{if\ } b_i=b_j \\
e^{\beta} & \text{otherwise}
\end{cases}
\end{equation}
We use here a blue line to indicate the presence of Pauli-$x$ (or $X$) at an edge in Eq. \eqref{eq:defectsIsing}. Its presence at an edge switches the interaction from ferromagnetic to anti-ferromagnetic while preserving the whole object as a valid partition function. Pauli-$x$ at an edge also denotes the symmetry action. It is possible to have a tensor network description of the partition function where all the local tensors are invariant under the action of $X$ on all the legs. It also implies that a partition function with a string of $X$'s will remain invariant under any continuous deformation in the string provided that the endpoints remain fixed.

At this point, it is instructive to write down an analytic expression for expectation value of average magnetization per two sites.
\begin{equation}
\begin{aligned}
\sum _{ \textbf{b} }^{  }{ \left( \cbox{3}{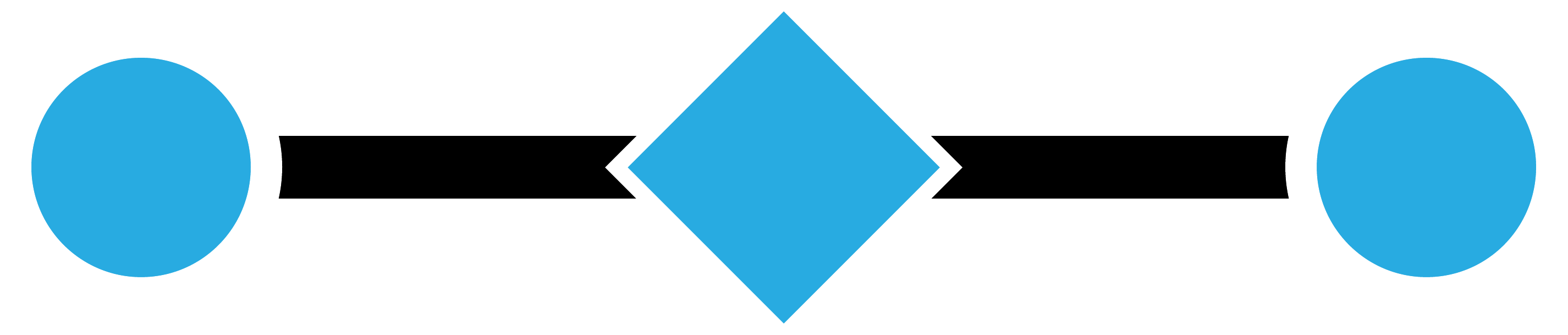} \right) \prod _{
	\substack{\left<i,j\right> \\ \neq \left<0,1\right>} }^{  }{ 
	\cbox{1.5}{figures/line_1}  }  } &= 
\sum _{ \textbf{s} }^{  }{ \left( s_0 + s_1 \right)e^{\beta} \prod _{
	\substack{\left<i,j\right> \\ \neq \left<0,1\right>} }^{  }{ e^{\beta s_i s_j} }
} \\ &=
{ \left( 1-\sinh ^{ -4 }{ \left( 2\beta  \right)  }  \right)  }^{ 1/8}
\end{aligned}
\label{eq:Z2}
\end{equation}
where the horizontal and vertical links in the product are expressed by an inclined edge. The cross sign on the edge indicates local order parameter $Z$ on the vertices labeled 0 and 1. Although, the link has negative weights, in analogy to Eq. \eqref{eq:defectsIsing}, we define it as follows.
\begin{equation}\label{eq:orderIsing}
\cbox{3}{figures/line_1x} :=
\begin{cases}
\left( -1\right)^{b_i} e^{\beta} & \text{if\ } b_i=b_j \\
0 & \text{otherwise}
\end{cases}
\end{equation}

\subsection{Anyonic vacuum and excitations}
An important object to inspect in order to analyze the norm of a quantum state is the on-site transfer operator. We start by writing it pictorially for \znqd{4} \qd\ with deformation.
\begin{equation}\label{eq:Etensor}
\cbox{3.5}{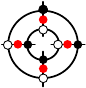} = \sum _{ k,l=0 }^{3}{ 
\cbox{2.6}{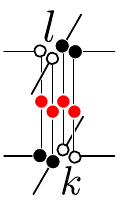}  } = \sum _{ k=0 }^{3}{ 
\cbox{2.6}{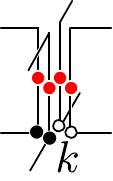}  }\ ,
\end{equation}
where each black circle in the sum with label $k$ denotes $X^{k}$ and $X$ is the generator of \z{4} with regular representation. The outline of circles specifies Hermitian conjugate. Red bubbles represent deformation $\exp\left(\theta X^2 \right)$. The last equality is possible since black and red circles commute and the on-site tensors are isometric. The deformation modeled by red bubble drive the system from \znqd{4} \qd\ to \zz{4}{2} \tc\ phase. Using Eq. \eqref{eq:Etensor} we can write the norm of vacuum as 
\begin{equation}
\left<0,1|0,1\right> = \sum _{ \substack{ k_i,l_i=0} }^{ 3 }{ \prod _{\substack{
		\text{rings} } }{  \cbox{3.5}{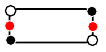} }  }  = \sum _{ \bm{t} }{
\prod _{\text{edges}}{  \cbox{1.5}{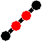} }  }\ ,
\label{eq:norm1}
\end{equation}
where the product is over all the rings (Fig. \ref{fig:correspond1}(b,c)) and by using Eq. \eqref{eq:Etensor} we can shrink each ring $\cbox{3.5}{figures/ring_1}$ to an edge $\cbox{1.5}{figures/line_3D}$ with the following definition.
\begin{equation}
\begin{aligned} 
\cbox{1.5}{figures/line_3D} &= 
\text{tr}\left( X^{ t_i }e^{ 2\theta X^{ 2 } }X^{ { t_j } } \right) \\&= 
\begin{cases}
2(e^{\theta}+e^{-\theta}) & \text{if } t_i - t_j=0 \ (\text{mod}\ 4) \\
2(e^{\theta}-e^{-\theta}) & \text{if } t_i - t_j=2 \ (\text{mod}\ 4) \\
0        & \text{otherwise}
\end{cases}
\end{aligned} 
\label{eq:boltzmanwt1}
\end{equation}
Each edge can be identified as an interaction in Ising model on square lattice. In order to be succinct we will write $\cbox{1.5}{figures/line_3D}$ as $\cbox{1.5}{figures/line_2}$. Since the Boltzmann weighs are zero if $ t_i-t_j =1$ (mod 2), we can write Eq. \eqref{eq:norm1} as a sum over two copies of Ising model on square lattice (Fig. \ref{fig:correspond1}b).
\begin{equation}
\left<0,1|0,1\right> = \sum_{t_i=0,2}{ \prod_{ \left<i,j \right> }{
	\cbox{1.5}{figures/line_2} } } + \sum_{t_i=1,3}{ \prod_{ \left<i,j \right> }{
	\cbox{1.5}{figures/line_2} } }
\label{eq:norm2}
\end{equation} 
Terms in Eq. \eqref{eq:Z} behave analogous to Eq. \eqref{eq:norm2}. By identifying different combinations in Eq. \eqref{eq:isingboltzmanwt1} and Eq. \eqref{eq:boltzmanwt1} with each other we can write
$$
{ e }^{ \beta  }=2\left( { e }^{ \theta  }+{ e }^{ -\theta  } \right)\ , \ 
{ e }^{ -\beta  }=2\left({ e }^{ \theta  }-{ e }^{ -\theta  } \right)
$$
which implies $\beta =\tanh ^{ -1 }{ { \left( { e }^{ -2\theta  } \right)  } }$.

Now, consider the anyon excitation $|0,i\rangle$ which gets confined as the system approaches the critical point. More precisely, the norm of $\left|0,i\right>$ is zero in the \zz{4}{2} \tc\ phase. Norm of the excitation, $\left<0,i|0,i\right>$, contains the ring   $\cbox{3.5}{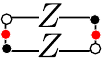} $ which shrinks to the edge $\dbox{5}{2.2}{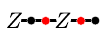}$.  We can write the norm as
\begin{equation}\label{eq:normexcit1}
\begin{aligned}
\left<0,i|0,i\right> & = 
\sum _{ \textbf{t} }{ 
\dbox{5}{2.2}{figures/line_3Dhxx}
\prod _{ \substack{\left<i,j\right> \\ \neq \left<0,1\right>} }^{  }{
	\cbox{1.5}{figures/line_3D}
}  } \\ &=
\sum _{ \textbf{t} }{ \left(\cbox{3}{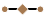} \right) \prod _{
		\substack{\left<i,j\right> \\ \neq \left<0,1\right>} }^{  }{  
		\cbox{1.5}{figures/line_2}  }  } 
\end{aligned}
\end{equation}
The brown diamond indicates a charge which is given by $Z:=Z_1$ in the ket and bra layer. Trace over each configuration on the edge is defined as follows
\begin{equation}
\begin{aligned} 
\cbox{3}{figures/line_2x} &=
\text{tr}\left( ZX^{ t_{ i } }e^{\theta X^{ 2 } }Ze^{\theta X^{ 2 } }X^{ t_j }
\right) \\&= 
\begin{cases}
2\left(\sqrt{-1}\right)^{t_i} & \text{if } t_i-t_j=0 \ (\text{mod}\ 4) \\
0        & \text{otherwise}
\end{cases}
\end{aligned} 
\label{eq:boltzmanwt2}
\end{equation}
We summarize the Boltzmann weight of all the configurations for two models in Tab.\ref{tab:boltzmann_comp}.

\begin{table*}[t]
	\centering
	\begin{tabular}{c|c|c|c || c|c|c|c}
		$\substack{ \text{Standard} \\ \text{Ising} \\ \text{model}}$ & 
		$\cbox{3.5}{figures/line_1h}$ & 
		$\cbox{3.5}{figures/line_1e}$ & 
		$\cbox{3.5}{figures/line_1x}$ & 
		$\cbox{3.5}{figures/line_2x}$ & 
		$\cbox{3.5}{figures/line_2e}$ & 
		$\cbox{3.5}{figures/line_2h}$ & 
		$\substack{ \text{Norm of} \\ \text{vacuum}}$ \\
		\hline\hline
		$b_i = b_j$ &
		$e^{\beta}$ &
		$e^{-\beta}$ &
		$\left( -1\right)^{b_i} e^{\beta}$ & 
		$2\left(\sqrt{-1}\right)^{t_i}$ & 
		$2(e^{\theta}-e^{-\theta})$ & 
		$2(e^{\theta}+e^{-\theta})$ & 		
		$t_i-t_j=0$ \ mod\ 4 	 \\
		\hline
		$b_i \neq b_j$ & 
		$e^{-\beta}$ & 
		$e^{\beta}$ & 
		$0$ & 
		$0$ & 
		$2(e^{\theta}+e^{-\theta})$ &
		$2(e^{\theta}-e^{-\theta})$ &		
		$t_i-t_j=2$ \ mod\ 4	 \\
		\hline
		\multicolumn{3}{c}{} & &
		$0$ & 
		$0$ & 
		$0$ & 
		$\begin{array}{c}
		t_i-t_j=1\ \text{mod}\ 4\\
		t_i-t_j=3\ \text{mod}\ 4
		\end{array}$
	\end{tabular}
	\caption{Comparison between the Boltzmann weights of standard Ising model(blue) and two decoupled copies of Ising models(brown) on square lattice which emerged from the norm of quantum vacuum.}
	\label{tab:boltzmann_comp}
\end{table*}

It is clear from the table that an edge $\cbox{3}{figures/line_2x}$ corresponds to the evaluation of magnetization per site up to a weighting factor $2(e^{\theta}+e^{-\theta})$. Using Eq. \eqref{eq:Z2} for magnetization per site and dividing by  $$2(e^{\theta}+e^{-\theta})  = \frac{1}{2} \left( { \tanh^{1/2} { \beta  }  }+{ 	\tanh^{-1/2} {\beta}} \right)$$ in order to compensate for the weighing factor we get an analytic expression for the norm of $\left|0,i\right>$. 
\begin{equation}\label{eq:0i0i}
\left< 0,i | 0,i \right> = \frac { 2{ \left( 1-\sinh ^{ -4 }{ 2\beta  }  \right)
	}^{ 1/8 } }{  { \tanh^{ 1/2 } { \beta  }  }+{ \tanh^{ -1/2 } {  \beta   }  }  }
\end{equation}

Excitation $\left|2,0\right>$ gets condensed to the vacuum in \zz{4}{2} \tc \ phase. Rings $\cbox{3.5}{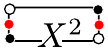}$ create a string of defective edges. In order to be consistent with notation used in Eq. \eqref{eq:defectsIsing}, we write $\cbox{3.5}{figures/ring_g2_g0}$ as $\cbox{3}{figures/line_2e}$. Tab.\ref{tab:boltzmann_comp} contains the Boltzmann weights for different configurations of $\cbox{3}{figures/line_2e}$ in column 6. Overlap of excitation $\left|2,1\right>$ with vacuum $\left|0,1\right>$ creates
a semi-infinite string of $\cbox{3}{figures/line_2e}$ edges (Fig. \ref{fig:correspond1}d). 
In order to get an analytic expression for condensate fraction $\left<0,1|2,1 \right>$, we first map the model from 2D classical Ising on square lattice to 1D quantum Ising chain. Kramers-Wannier duality of 2D classical Ising manifest itself as 1D quantum Ising duality using disorder operators on dual lattice.
\begin{equation}\label{eq:disorderops}
\begin{split}
\tau_{i+1/2}^{z}=\prod_{j \leq i}{\sigma_{j}^{x}}
\end{split}
\quad,\quad
\begin{split}
\tau_{i+1/2}^{x}={\sigma_{i}^{z}}{\sigma_{i+1}^{z}}
\end{split}
\end{equation}
$\sigma_i^x \left(\tau^x_{i+1/2}\right)$ and $\sigma_i^z \left(\tau^z_{i+1/2}\right)$ are Pauli matrices on (dual) square lattice. Using this transformation, a semi-infinite domain wall created by a string of $X$'s $(...XXX)$ translates into point operator corresponding to magnetization per site on the dual lattice. Condensate fraction $\left<0,1|2,1 \right>$ can be
written analytically as
\begin{equation}\label{eq:2101}
\left< 2,1 | 0,1 \right> = { \left( 1-\sinh ^{ -4 }{ 2\beta^* }  \right)  }^{
	1/8 }, 
\end{equation}
where $\beta^*$ is related to $\beta$ by $\sinh 2\beta \ \sinh 2\beta^*=1$.

We have computed analytically two condensate fractions in Eq. \eqref{eq:0i0i} and Eq. \eqref{eq:2101}. The rest of non-zero but not constant overlaps can be proved equal to either of the two using following identities.
\begin{equation}\label{eq:identities1}
\cbox{3.5}{figures/ring_1} =
\cbox{3.5}{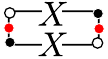}=
\cbox{3.5}{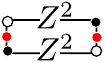}
\end{equation}
\begin{equation}\label{eq:identities2}
\cbox{3.5}{figures/ring_x_x} =
\cbox{3.5}{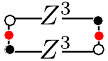}
\end{equation}
The equality holds for every value of $\theta$ along the phase transition. Analytic results conforms exactly with the numerical data in Fig. \ref{fig:QD_TC_Ising}a.

The norm of vacuum for the phase transition between \znqd{4} \qd\ and \zz{2}{1} \tc\ with explicit symmetry breaking can also be mapped to classical Ising model where the Ising variables have different Boltzmann weights. \znqd{4} \qd\ $\leftrightarrow$ \zz{4}{2} \ds\ phase transition also gives rise to an Ising model but the identification of Ising variables is subtle as the local tensor network description of \ds\ model breaks the rotational symmetry.

\subsection{Simplifications by explicit symmetry breaking}

We conclude this appendix with a note on how explicit symmetry breaking
can prove useful in simplifying the mapping between certain interpolations
and classical models (not necessarily restricted to the Ising model). To
this end, let us consider the interpolations between DS and \zz42 TC and
\zz21 obtained by direct interpolation of the on-site transfer operators
$\mathbb E$.  In this case, it is straightforward to check that for all
three on-site transfer operators, it holds that
$\mathbb E = \mathbb E\,P_0^{\otimes 4} + \mathbb E\,P_1^{\otimes 4}$,
where each $P_c$ acts on a ket-bra pair of indices, and $P_c$ projects
onto the 4-dimensional space of ket-bra operators spanned by
$X^{2a+c}Z^{2b}$. This is, $\mathbb E$ has a \emph{fully local} block
structure corresponding to the blocks $P_c$, $c=0,1$, which relates to the
fact that all tensors in the family break the same $\mathbb Z_2$ symmetry
(namely \zz44 to \zz42). By projecting $\mathbb E$ locally onto, e.g.,
$P_0$, we can break this symmetry explicitly, and thereby replace the
corresponding interpolations with interpolations with a (ket+bra) bond dimension
of $4$. This, on the one hand, facilitates possible mappings to classical
models (such as a mapping of the DS to \zz42 TC interpolation to a loop
model, and of the DS to \zz42 interpolation to a classical transfer
operator with breaking of a $\mathbb Z_2\times \mathbb Z_2$ symmetry), and
at the same time, it allows for more efficient numerical simulations, used
e.g.\ for the large $\chi$ data in Fig.~\ref{fig:compare_corrs}a.

\section{Unitary equivalence between \zz{4}{2} \ds\ and \zz{2}{1} \tc}\label{app:z2-dsmapping}
In this appendix, we give an MPO construction of a unitary which transform the transfer operators of \zz{4}{2} \ds\ and \zz{2}{1} \tc\ into each other. It is helpful for later purposes to first write down the local tensors which represent the RG fixed points of two phases. In the case of \zz{4}{2} \ds\ model
\begin{equation}\label{eq:DSNotation}
\cbox{3}{figures/ds_mpo}\ ,\ 
\cbox{3}{figures/ds_tensor_1}=
\begin{pmatrix}
\{\mathds{1}\}_{a=0,b=0} & \{X^2 Z^2\}_{a=0,b=1} \\
\{Z^2\}_{a=1;b=0} & \{X^2\}_{a=1;b=1}
\end{pmatrix}
\end{equation}
Green ring is an MPO projector for \zz{4}{2} \ds\ applied on \znqd{4} \qd\ (black ring). In later usage, we will drop the subscripts in matrix notation. Similarly, for \zz{2}{1} \tc
\begin{equation}\label{eq:Z2TCNotation}
\cbox{3}{figures/Z2TC_mpo}\ ,\ 
\cbox{3}{figures/Z2TC_tensor}=
\begin{pmatrix}
\mathds{1}+Z^2 & 0 \\
0 & \mathds{1}-Z^2
\end{pmatrix}
\end{equation}
Moreover, the on-site transfer operators (termed ``double tensors'' in the following)
of two models have the same representation as given in Eq. \eqref{eq:DSNotation} and Eq. \eqref{eq:Z2TCNotation}. The transfer operator constructed by blocking the double tensors, $\mathbb{E}(\theta) = \theta \mathbb{E}_{\text{DS}} + (1-\theta) \mathbb{E}_{\text{TC}}$, can be interpreted as a sum of alphabetic strings where each alphabet is either \ds\ or \tc\ double tensor. And the local tensor $u$ of the desired MPO unitary $U$ is expected to swap the double tensor of \zz{4}{2} \ds\ with the double tensor of \zz{2}{1} \tc. Local
tensor $u$ should act to substitute the double tensor of \ds\ with the double tensor of \tc\ and vice-versa.

Motivated by the construction of discriminating string order parameters for topological phases in
\cite{duivenvoorden2013symmetry}, we start by writing the local tensor description of MPO $U$. 
\begin{equation}\label{eq:dstc_u}
\cbox{3.5}{figures/mpotensor_U}=
\begin{pmatrix} 
\mathds{1}\otimes\left(\mathds{1}+X^2\right) & \mathds{1}\otimes\left(\mathds{1}-X^2\right) \\
X\otimes\left(X^3-X\right)&  X\otimes\left(X^3+X\right)  \end
{pmatrix},
\end{equation}
where the top (bottom) index is identified with the row (column) index of the matrix and the arrow head points in the direction of column index. Now, we show why Eq. \eqref{eq:dstc_u} is the right description of $u$ by showing its action on the on-site transfer operators. With $u$ defined in Eq. \eqref{eq:dstc_u}, its action on \zz{4}{2} \ds\ tensor is
\begin{equation}\label{eq:u_on_ds1}
\cbox{3.0}{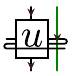}=
\begin{pmatrix}
X_+ & Z^2X_+ & X_-& -Z^2X_- \\
Z^2X_+ & X_+ & Z^2X_- & -X_- \\ 
X_- & Z^2X_- & X_+ & -Z^2X_+ \\
-Z^2X_- & -X_- & -Z^2X_+ & X_+
\end{pmatrix},
\end{equation}
where $X_{\pm}=\mathds{1} \pm X^2$. Although, it is not very clear in above form, it is more insightful to understand the action by a unitary transformation. Consider a unitary $M$ with the following definition, 
\begin{equation}
\cbox{3}{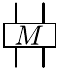}:=\frac{1}{2}
\begin{pmatrix}
-1 & 1 & 1 & 1 \\
1 & -1 & 1 & 1 \\
1 & 1 & -1 & 1 \\
1 & 1 & 1 & -1 \\
\end{pmatrix}
\end{equation}
where the join of top(bottom) indices correspond to row(column) index of the matrix. By applying $M$ to Eq. \eqref{eq:u_on_ds1}, we obtain
\begin{equation}\label{eq:A_on_UDS}
\cbox{2.6}{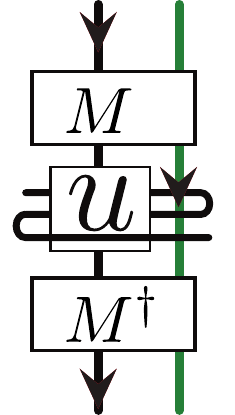}=
\begin{pmatrix}
Z_+ & 0 & 0 & 0 \\
0 & Z_+X^2 & 0 & 0 \\
0 & 0 & Z_- & 0 \\
0 & 0 & 0 & Z_- X^2 \\
\end{pmatrix}
\end{equation}
where $Z_{\pm}=\mathds{1}\pm Z^2$. Matrix entries across the main diagonal correspond to the four blocks (or fixed points since each block can be identified with a fixed point) of \zz{2}{1} \tc\ (see Eq. \eqref{eq:z2tc-fxpoints}). MPO projectors (blue and green rings) in Eq. \eqref{eq:DSNotation} and Eq. \eqref{eq:Z2TCNotation} commutes with the black MPO of \znqd{4} \qd, so the action of $u$ on the \ds\ tensor can be summarized as
\begin{equation}\label{eq:uds-z2tc}
\cbox{4}{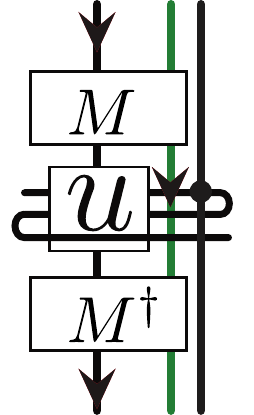}=\cbox{4}{figures/Z2TC_UEffect}\, 
\end{equation}
where,
\begin{equation}
\cbox{2}{figures/Z2TC_UEffect_Black}=
\begin{pmatrix}
\mathds{1} & 0& 0& 0\\
0 & X& 0& 0 \\
0 & 0& X^2& 0 \\
0 & 0& 0& X^3
\end{pmatrix}.
\end{equation}
Now, we consider the action of $u$ on the local tensor of \zz{2}{1} \tc.
\begin{equation}\label{eq:u_on_z2tc}
\cbox{3}{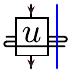}=
\begin{pmatrix}
X_+ Z_+ & X_- Z_+ & 0 & 0\\
X_- Z_- & X_+ Z_- & 0 & 0\\
0 & 0 & X_+ Z_- & X_- Z_-\\
0 & 0 & X_- Z_+ & X_+ Z_+
\end{pmatrix}
\end{equation}
In order to study the structure of \zz{4}{2} \ds\ blocks, again we define a unitary
\begin{equation}
\cbox{3}{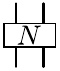}:=\frac{1}{\sqrt{2}}
\begin{pmatrix}
1 & 1 & 0 & 0 \\
1 & -1 & 0 & 0 \\
0 & 0 & -1 & 1 \\
0 & 0 & 1 & 1 \\
\end{pmatrix}
\end{equation}
By doing a unitary transformation on Eq. \eqref{eq:u_on_z2tc},
\begin{equation}
\cbox{3.0}{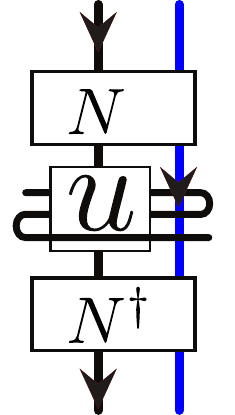}=
\begin{pmatrix}
\mathds{1} & X^2Z^2 & 0 & 0\\
Z^2 & X^2 & 0 & 0 \\
0 & 0 & \mathds{1} & X^2Z^2 \\
0 & 0 & Z^2 & X^2 
\end{pmatrix}
\end{equation}
The two blocks are completely identical and correspond to one of the symmetry broken fixed point of \zz{4}{2} \ds\ model. Similar to Eq. \eqref{eq:uds-z2tc}, the action of $u$ on \zz{2}{1} tensor with the local tensor of \znqd{4} \qd\ produces the local tensor of \zz{4}{2} \ds\ model.
\begin{equation}\label{eq:uz2tc-ds}
\cbox{4}{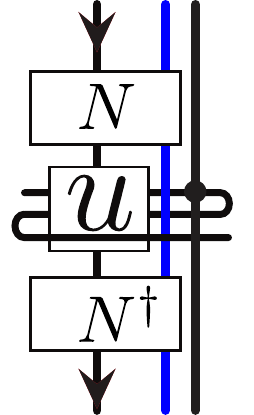}=\cbox{4}{figures/DS_UEffect}
\end{equation}
Furthermore, from the action of $u$ in Eq. \eqref{eq:uds-z2tc} and Eq. \eqref{eq:uz2tc-ds}, we can also verify that the following relation also holds between $u$ and the on-site transfer operators of \zz{4}{2} \ds\ and \zz{2}{1} \tc.
\begin{equation}\label{eq:dstcLocal}
\begin{aligned}
\cbox{6.5}{figures/duality_1Left}\ &= \cbox{6.5}{figures/duality_1Right}\ , \\
\cbox{6.5}{figures/duality_2Left}\ &= \cbox{6.5}{figures/duality_2Right}\ ,
\end{aligned}
\end{equation}
where,
\begin{equation}
\cbox{3}{figures/blocked_DS} = \cbox{3}{figures/ds_mpo}\ , \ 
\cbox{3}{figures/blocked_Z2TC} = \cbox{3}{figures/Z2TC_mpo}
\end{equation}
In order to obtain the equation $U \mathbb T(\theta) U^\dagger = \mathbb
T(1-\theta)$ of Sec.~\ref{subsec:tc21-to-ds}, we insert $u$ at one
end of the transfer operator and by using Eq.~\eqref{eq:dstcLocal} and by
zipping $u$ to the other end of the transfer operator we can achieve the
global action of $U$ as required.

\end{document}